\documentclass[11pt]{article}
\usepackage{url}

\usepackage{hyperref}

\usepackage{amssymb,amscd,amstext,amsmath, amsthm, graphicx}
\usepackage{latexsym}
\usepackage[russian,english]{babel}
\usepackage{floatflt}
\usepackage[mathscr]{eucal}
\usepackage{tikz}
\usetikzlibrary{arrows}
\usepackage{verbatim}
\usepackage{soul}
\usepackage{wrapfig}
\usepackage{fancyhdr}
\usepackage{hyperref}
\usepackage{setspace}
\usepackage{mathdots}
\usepackage{blindtext}
\usepackage{enumitem}
 \usepackage{placeins}
 \usepackage{algorithm}
 \usepackage{csquotes}
\usepackage{multirow}
\usepackage{longtable}

\let\OLDthebibliography\thebibliography
\renewcommand\thebibliography[1]{
  \OLDthebibliography{#1}
  \setlength{\parskip}{0pt}
  \setlength{\itemsep}{0pt plus 0.3ex}
}

\usepackage[top=1in, bottom=1in, left=1in, right=1in]{geometry}

\newcommand{\ket}[1]{\ensuremath{\left|#1\right\rangle}}

\usepackage{tikz}\usetikzlibrary{shadows}

\usetikzlibrary{shapes.geometric, arrows}

\tikzstyle{box} = [rectangle, rounded corners, minimum width=1cm, minimum height=1cm,text centered, draw=black, fill=yellow!30]
\tikzset{oplus/.style={path picture={%
    \draw[black]
    (path picture bounding box.south) -- (path picture bounding box.north)
    (path picture bounding box.west) -- (path picture bounding box.east);
}}}
\tikzstyle{xor} = [oplus, draw = black, circle, minimum size = 0.5cm]

\newcommand{\T}{\top}

\newcommand\W{6560}
\newcommand\w{1312}
\newcommand\K{20}
\newcommand\Kplus{21}
\newcommand\Ktwo{40}
\newcommand\F{\mathbb{F}}

\usetikzlibrary{quantikz}

\newsavebox{\boxX}
\newsavebox{\boxZ}
\newsavebox{\boxH}
\newsavebox{\boxCnot}
\newsavebox{\boxSwap}
\newsavebox{\boxToffoli}

\theoremstyle{definition}
\newtheorem{remark}{Remark}

\definecolor{linkcolor}{HTML}{000000}
\definecolor{urlcolor}{HTML}{000000}
\hypersetup{pdfstartview=Fit, linkcolor=linkcolor,urlcolor=urlcolor,
colorlinks=true, citecolor=linkcolor}
\renewenvironment{abstract}{%
\begin{center}\begin{minipage}{0.9\textwidth}\begin{small}
\textbf{Abstract.}}
{\end{small}\par\noindent\end{minipage}\end{center}}

\title{\bf An overview of the Eight International Olympiad in Cryptography ``Non-Stop University CRYPTO'' \footnote{The work of the first, second, ninth, tenth, eleventh authors was carried out within the framework of the state contract of the Sobolev Institute of Mathematics (project no. FWNF-2022-0018).}
}

\author{A.~Gorodilova$^{1}$,
    N.~Tokareva$^{1}$,
    S.~Agievich$^{2}$,
    I.~Beterov$^{3,4}$,
    T.~Beyne$^{5}$,
    L.~Budaghyan$^{6}$,\\
    C.~Carlet$^{6,7}$,
    S.~Dhooghe$^{5}$,
    V.~Idrisova$^{1}$,
    N.~Kolomeec$^{1}$,
    A.~Kutsenko$^{1}$,
    E.~Malygina$^{8}$,\\
    N.~Mouha$^{9}$,
    M.~Pudovkina$^{10}$,
    F.~Sica$^{11}$,
    A.~Udovenko$^{12}$
    \\
  \\
  {\small$^{1}$Sobolev Institute of Mathematics, Novosibirsk, Russia} \\
  {\small$^{2}$Belarusian State University, Minsk, Belarus} \\
  {\small$^{3}$Rzhanov Institute of Semiconductor Physics SB RAS, Novosibirsk, Russia} \\
  {\small$^{4}$Novosibirsk State University, Novosibirsk, Russia} \\
  {\small$^{5}$imec-COSIC, ESAT, KU Leuven, Belgium}\\
  {\small$^{6}$University of Bergen, Bergen, Norway}\\
  {\small$^{7}$University of Paris 8, Paris, France}\\
  {\small$^{8}$Immanuel Kant Baltic Federal University, Kaliningrad, Russia}\\
  {\small$^{9}$Strativia, Largo, United States}\\
  {\small$^{10}$National Research Nuclear University MEPhI, Moscow, Russia}\\
  {\small$^{11}$Nazarbayev University, Nur-Sultan, Kazakhstan}\\
  {\small$^{12}$CryptoExperts, Paris, France}\\
    \\
    {\small E-mail: {\tt nsucrypto@nsu.ru}}
    }

\date{}

\begin{document}

\hypersetup{pageanchor=false}

\begin{titlepage}
\maketitle
\begin{abstract}
Non-Stop University CRYPTO is the International Olympiad in Cryptography that was held for the eight time in 2021. Hundreds of university and school students, professionals from 33 countries worked on mathematical problems in cryptography during a week. The aim of the Olympiad is to attract attention to curious and even open scientific problems of modern cryptography. In this paper, problems and their solutions of the Olympiad'2021 are presented. We consider 19 problems of varying difficulty and topics: ciphers, online machines, passwords, binary strings, permutations, quantum circuits, historical ciphers, elliptic curves, masking, implementation on a chip, etc. We discuss several open problems on quantum error correction, finding special permutations and s-Boolean sharing of a function, obtaining new bounds on the distance to affine vectorial functions.

\vspace{0.2cm}

\end{abstract}
\end{titlepage}

\hypersetup{pageanchor=true}
\pagenumbering{arabic}

\setcounter{page}{2}

\section{Introduction}

{\bf Non-Stop University CRYPTO} ({\bf NSUCRYPTO}) \cite{nsucrypto} is the unique international cryptographic Olympiad in the world. It contains scientific mathematical problems for professionals, school and university students. Its aim is to involve young researchers in solving curious and tough scientific problems of modern cryptography. From the very beginning, the concept of the Olympiad was not to focus on solving olympic tasks but on including unsolved research problems at the intersection of mathematics and cryptography. Everybody can participate the Olympiad as far as it holds via the Internet. Rules and format of the Olympiad can be found at the official website \cite{nsucrypto-rules}. 

Non-Stop University CRYPTO history started in 2014. We were inspired by an experience of the Russian Olympiad in Mathematics and Cryptography for school-students and decided to organize an International event with real scientific content for students and professionals. Since then eight Olympiads were held and more than 3000 students and specialists from 68 countries took part in it. The Program committee consists of 31 members from cryptographic groups all over the world. Between them are creators of several modern technologies and ciphers, like AES, Chaskey, etc. Main organizers are Cryptographic center (Novosibirsk), Mathematical Center in Akademgorodok, Novosibirsk State University, Sobolev Institute of Mathematics, KU Leuven, Belarusian State University, Tomsk State University and Kovalevskaya North-West Centre of Mathematical~Research.


In 2021, the Olympiad was dedicated to the 100th anniversary of the Cryptographic Service of Russian Federation. There were 746 participants from 33 countries; 32 participants in the first round and 40 teams in the second round from 21 countries became the winners (see the list \cite{nsucrypto-winners}). 19 problems were proposed to participants and 4 of them included open questions. 

According to the results of each Olympiad, scientific articles are published with an analysis of the solutions proposed to the participants, including unsolved ones, see \cite{nsucrypto-2015, nsucrypto-2014,  nsucrypto-2017, nsucrypto-2018, nsucrypto-2019, nsucrypto-2020, nsucrypto-2016}.

\section{An overview of open problems}

A specialty of the Olympiad is that unsolved problems at the intersection of mathematics and cryptography are formulated to the participants along with problems with known solutions. 
All the open problems stated during the Olympiad history as well as their current status can be found at the Olympiad website \cite{nsucrypto-unsolved}. There are 26 open problems in this list.

The variety and difficulty of the problems are wide. In fact, we suggest problems that are of great interest to cryptography over which many mathematicians are struggling in search of a solution. For example, these problems include ``APN permutation'' (2014), ``Big Fermat numbers'' (2016), ``Boolean hidden shift and quantum computings'' (2017), ``Disjunct Matrices'' (2018), and others. For instance, the problem ``8-bit S-box'' (2019) was inspired by \cite{19-Fomin}.

Despite the fact that hard problems can be found in the list of the Olympiad problems, participants are not afraid to take on such tasks. Indeed, some of the problems we suggested can be solved or partially solved even during the Olympiad. For example, the problems ``Algebraic immunity'' (2015), ``Sylvester matrices'' (2018), ``Miller~---~Rabin revisited'' (2020) were solved completely. 
Also, partial solutions were suggested for the problems ``Curl27'' (2019), ``Bases'' (2020),  ``Quantum error correction'' (2021, see section\;\ref{sec-quantum}) and ``s-Boolean sharing'' (2021, see section\;\ref{sec-sharing}). 

Furthermore, some researchers are working on finding solutions after the Olympiad was over. In \cite{20-Kiss-Nagy}, a complete solution was found for the problem ``Orthogonal arrays'' (2018). The authors have shown that no orthogonal arrays $OA(16\lambda,11,2,4)$ exist with $\lambda=6$ and $7$. 
Another problem, ``A secret sharing'' (2014) was partially solved in \cite{17-Geut}, \cite{19-Geut}, where particular cases were considered, and was recursively solved in \cite{19-Ayat}.

\section{Problem structure of the Olympiad}
\label{problem-structure}

There were 19 problems stated during the Olympiad, some of them were included in both rounds (Tables\;\ref{Probl-First},\,\ref{Probl-Second}). Sections A, B of the first round each consisted of seven problems. The second round was composed of ten problems; four of them included unsolved questions (awarded special prizes). 

\begin{table}[H]
\centering\footnotesize
\caption{{\bf Problems of the first round}}
\medskip
\label{Probl-First}
\begin{tabular}{cc}
\begin{tabular}{|c|l|c|}
  \hline
  N & Problem title & Maximum scores \\
  \hline
  \hline
  1 & \hyperlink{pr-sheet}{Have a look and read!} & 4 \\
    \hline
  2 & \hyperlink{pr-2021}{2021-bit key}  & 4 \\
    \hline
  3 &  \hyperlink{pr-con}{A conundrum} & 4\\
    \hline
  4 &  \hyperlink{pr-pass}{Related passwords} & 4 \\
  \hline
    5 & \hyperlink{pr-planet}{A space message} & 4 \\
 \hline
  6 &  \hyperlink{pr-seq}{Two strings} & 4\\
    \hline
  7 & \hyperlink{pr-present}{A small present for you!} & 4 \\
 \hline

\end{tabular}

&

\begin{tabular}{|c|l|c|}
  \hline
  N & Problem title & Maximum scores \\
  \hline
    \hline
  1 & \hyperlink{pr-sheet}{Have a look and read!} & 4 \\
  \hline
 2 &  \hyperlink{pr-seq}{Two strings} & 4\\
   \hline
   3 & \hyperlink{pr-planet}{A space message} & 4 \\
 \hline
 4    & \hyperlink{pr-residue}{Elliptic curve points} & 4 \\
  \hline 
  5 & \hyperlink{pr-Nrounds}{The number of rounds} & 6 \\
    \hline
  6 & \hyperlink{pr-present}{A present for you!} & 6 \\
   \hline
  7 & \hyperlink{pr-ent-st}{Try your quantum skills!} & 6 \\
    \hline
\end{tabular}
\\
{\bf Section A}   &  {\bf Section B}\\
\end{tabular}
\end{table}

\vspace{-0.6cm}

\begin{table}[H]
\centering\footnotesize
\caption{{\bf Problems of the second round}}
\medskip
\label{Probl-Second}
\begin{tabular}{|c|l|c|}
  \hline
  N & Problem title & Maximum scores \\
  \hline
    \hline
  1 &  \hyperlink{pr-con}{A conundrum} & 4\\
    \hline
  2 &  \hyperlink{pr-APN+lin-perm}{Let's find permutations!} & open problem\\
    \hline
  3 &  \hyperlink{pr-ballots}{Shuffle ballots} & 8 \\
    \hline
  4 &  \hyperlink{pr-decode}{Let's decode!} & 6 \\
    \hline
  5 &  \hyperlink{pr-hiding}{Nonlinear hiding} & 7\\
    \hline
  6 &  \hyperlink{pr-feistel}{Studying Feistel schemes} &  10\\
    \hline
  7 &\hyperlink{pr-sB-sharing}{$s$-Boolean sharing}  & open problem\\
    \hline
  8 & \hyperlink{pr-quantum}{Quantum error correction} & open problem\\
    \hline
 9 & \hyperlink{pr-2021}{2021-bit key}  & 4 \\
    \hline
  10 & \hyperlink{pr-xorprus}{Close to permutations}  & 8 \\
     \hline
  11 &  \hyperlink{pr-distance}{Distance to affine functions} & open problem\\
   \hline
  12 & \hyperlink{pr-Nrounds}{The number of rounds} & 6 \\
    \hline
  13 & \hyperlink{pr-present}{A present for you!} & 6 \\
   \hline

\end{tabular}
\end{table}

\section{Problems and their solutions}\label{problems}

In this section, we formulate all the problems of 2021 year Olympiad and present their detailed solutions paying attention to solutions proposed by the participants.

\subsection{Problem ``Have a look and read!''}

\subsubsection{Formulation}
\hypertarget{pr-sheet}{}

\begin{figure}
\centering
\begin{tabular}{ccc}
\includegraphics[width=0.35\textwidth]{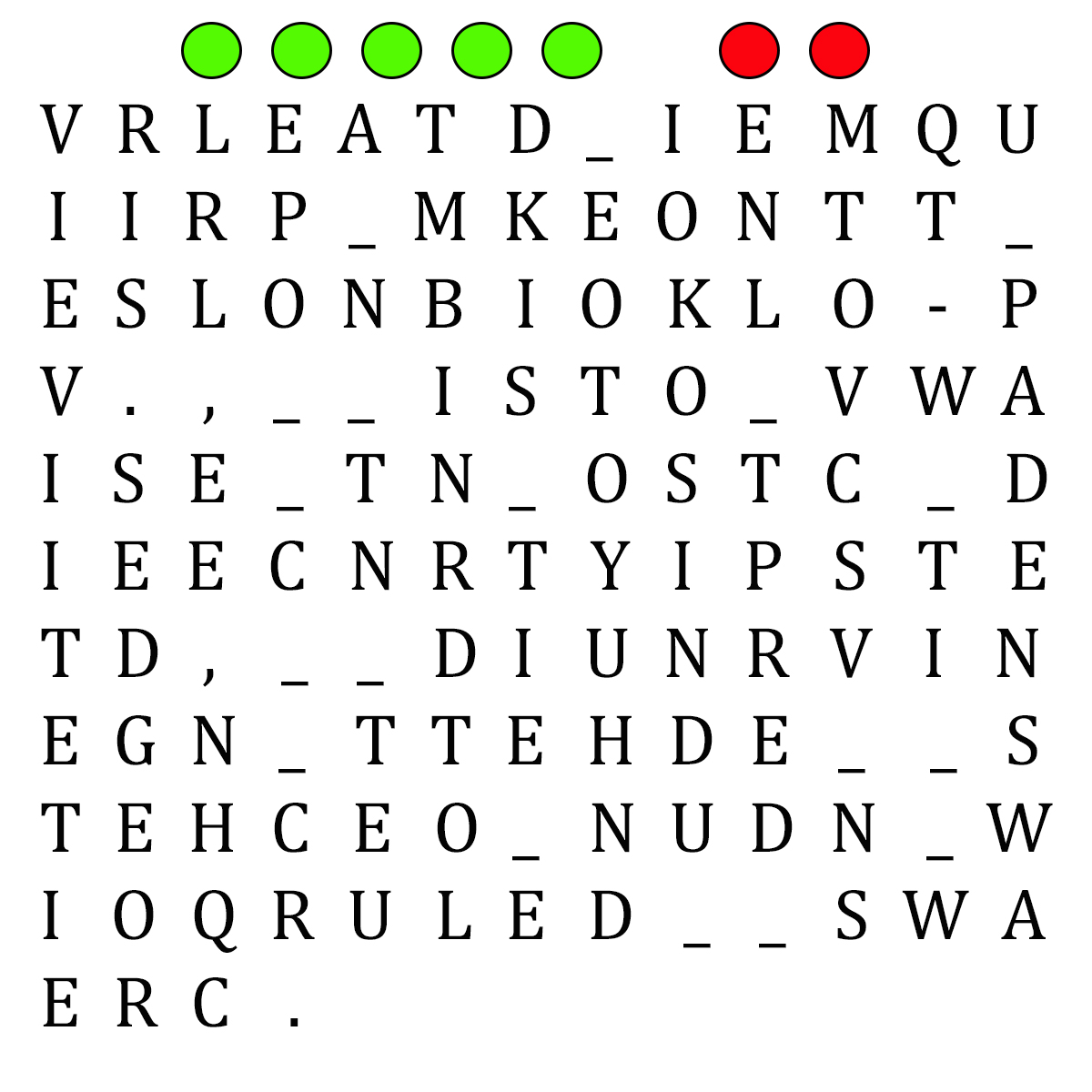}
& ~~~~~~ &
\includegraphics[width=0.35\textwidth]{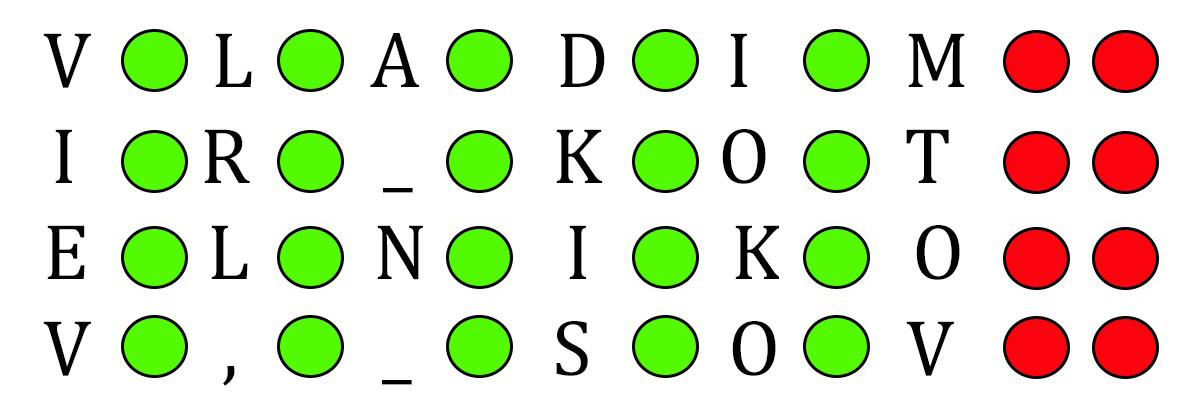}
\\
(a) Formulation.
& ~~~~~~ &
(b) Solution.
\end{tabular}
\caption{Illustrations for the problem ``Have a look and read!''.}
\label{fig:Kot}
\end{figure}

Read a secrete message in Fig.~\ref{fig:Kot}(a).


\subsubsection{Solution}

This is a permutation cipher. The circles above the text are hints how to read a message. Fig.~\ref{fig:Kot}(b) illustrates how to read the beginning of the message. The rest lines can be read the same way. 

After that, the rest letters can be read and form the whole message.
The answer is ``Vladimir Kotelnikov, Soviet scientist, invented the unique secret equipment SOBOL-P. It was not decrypted during the Second World War''.

\subsection{Problem ``2021-bit key''}

\subsubsection{Formulation}
\hypertarget{pr-2021}{}

A pseudo-random generator produces sequences of bits (that is of $0$ and $1$) step by step. To start the generator, one needs to pay 1 {\sl nsucoin} and the generator produces a random bit (that is a sequence of length $1$). Then, given a generated sequence $S$ of length~$\ell$, $\ell\geqslant1$, one of the following operations can be applied on each step:

\begin{enumerate}[noitemsep]
\item[{\bf 1.}] A random sequence of 4 bits is added to $S$, so a new sequence $S'$ has length $\ell+4$.
The charge for using this operation is 2 {\sl nsucoins}.
\item[{\bf 2.}] A random sequence of $2\ell$ bits is added to $S$, so a new sequence $S'$ has length $3\ell$.\\
The charge for using this operation is 5 {\sl nsucoins}.
\end{enumerate}

Bob needs to generate a secret key of length exactly 2021 bits for his new cipher. What is the minimal number of {\sl nsucoins} that he has to pay for the key?

\subsubsection{Solution}

First of all, it should be noted that as soon as a multiplication by 3 appears in the sequence of actions, then further after it, no more than two additions of 4 can be used. Indeed, if we assume three additions, we have 
$\ell \rightarrow 3\ell+3*4$ that equals $3*(\ell+4)$ but is more expensive. 
 Therefore, to minimize the cost, the sequence has the form $((1 + 4 + 4 + \ldots + 4) * 3 \ldots)$, where after the first multiplication by 3 there are no more than two additions of 4 in a row.

Now let us find the sequence of actions starting from the end. If the length is not divisible by~3, then it is necessary to subtract 4. If it is divisible by 3, then it is necessary to check what is cheaper: to divide by 3 or to fill this piece completely by 4s.

Thus, we come to the following sequence of actions:
\begin{center}
\begin{tabular}{|c|l||c|l||c|l}
    1 & $2021-4=2017$   & 5 & $667-4=663$ & 9  & $213:3=71$\\
    2 & $2017-4=2013$   & 6 & $663:3=221$ & 10 & $71-4=67$\\
    3 & $2013:3=671$    & 7 & $221-4=217$ & 11 & $67-4=63$\\
    4 & $671-4=667$     & 8 & $217-4=213$ & 12 & $63:3=21$\\
\end{tabular}
\end{center}

And here we can see that to get 21 by 4s is cheaper than multiplying by 3. Therefore, the last steps all consist of subtracting fours.

Thus, at least 47 {\sl nsucoins} is required to get a sequence of 2021 length.

\subsection{Problem ``A conundrum''}

\subsubsection{Formulation}
\hypertarget{pr-con}{}

Here is the conundrum sent by Alice to Bob:
\begin{quote}
{\tt b tn ztwobfc twxfc t hutek vptbwbfc t svbeo hukbfq nu vx ntpo xlv \\wbfus b ztwo 6 nuvpus fxpvk vkuf 2 nuvpus zusv 5 nuvpus utsv 6 nuvpus \\ sxlvk 3 nuvpus utsv 6 nuvpus fxpvk 4 nuvpus utsv 3 nuvpus sxlvk \\3 nuvpus zusv 3 nuvpus fxpvk 6 nuvpus sxlvk 3 nuvpus fxpvk 4.24 nuvpus\\ sxlvkutsv 3 nuvpus utsv 1 nuvpu zusv 6 nuvpus fxpvk 1 nuvpu zusv \\3 nuvpus utsv 6 nuvpus sxlvk 6 nuvpus fxpvk 6.49 nuvpus sxlvksxlvkutsv \\6 nuvpus fxpvk 4 nuvpus utsv 3 nuvpus zusv 6 nuvpus sxlvk 3 nuvpus utsv \\3 nuvpus fxpvk tfq 1 nuvpu zusv zktv bs vku ftnu vktv b ktau zpbvvuf bf \\vku stfq?}
\end{quote}
Find an answer to Alice's question!

\subsubsection{Solution}

This is a classic substitution cipher: each letter in the cipher represents another in the plaintext. A good place to start decoding is the single letter words ``{\tt b}'' and ``{\tt t}'', which must correspond to the single letter words ``a'' and ``I'' in English, though we don't immediately know which way round. Examination of two and three letter words suggests that ``{\tt vku}'' likely represents ``the'', and so on. With a bit of experimentation, we find the plaintext:
\begin{quote}
    ``I am walking along a beach with a stick to mark out lines on the sand. I walk 6 metres north, then 2 metres west, 5 metres east, 6 metres south, 3 metres east, 6 metres north, 4 metres east, 3 metres south, 3 metres west, 3 metres north, 6 metres south, 3 metres north, 4.24 metres southeast, 3 metres east, 1 metre west, 6 metres north, 1 metre west, 3 metres east, 6 metres south, 6 metres north, 6.49 metres southsoutheast, 6 metres north, 4 metres east, 3 metres west, 6 metres south, 3 metres east, 3 metres north and 1 metre west. What is the name that I have written in the sand?''
\end{quote}
Following the instructions, we trace out the letters: {\tt TURING}.

\subsection{Problem ``Related passwords''}

\subsubsection{Formulation}
\hypertarget{pr-pass}{}

Tim and Ann want to create curiously related passwords for their cryptosystem. 
A password is a 9-digit decimal number.
To start, they choose a random number $e_1e_2...e_9$ that has nine (not necessarily distinct) decimal digits.
\begin{itemize}[noitemsep]
\item Tim finds a password $d_1d_2...d_9$ such that each of the numbers formed by replacing just one of the digits $d_i$ in $d_1d_2...d_9$ by the corresponding digit $e_i$  is divisible by 7.

\item Ann finds a password $f_1f_2...f_9$ in similar but not the same way: each of the nine numbers formed by replacing one of the $e_i$ in $e_1e_2...e_9$ by $f_i$ is divisible by 7.
\end{itemize}

Show that for each $i$, $d_i - f_i$ is divisible by 7 for any of Tim's and Ann's passwords!


\subsubsection{Solution}

Let us denote $D = d_1d_2...d_9$ and $E = e_1e_2...e_9$.
Since $(e_i - d_i)10^{9-i} + D = 0 \pmod{7}$ for $i=1,...,9$, then summing up all these equalities we get $E - D + 9D = 0 \pmod{7}$. Hence, $E+D$ is divisible by 7.
Also, we have that $(f_i - e_i)10^{9-i} + E = 0 \pmod{7}$ for $i=1,...,9$. Therefore, $(f_i - d_i)10^{9-i} + D + E = 0 \pmod{7}$ for any $i$. Since 10 is coprime with 7 and 7 divides $E + D$, we get that $d_i - f_i$ is divisible by 7 for any $i$.

\subsection{Problem ``A space message''}

\subsubsection{Formulation}
\hypertarget{pr-planet}{}

\begin{figure}[h!]
\centering
\includegraphics[width=0.35\textwidth]{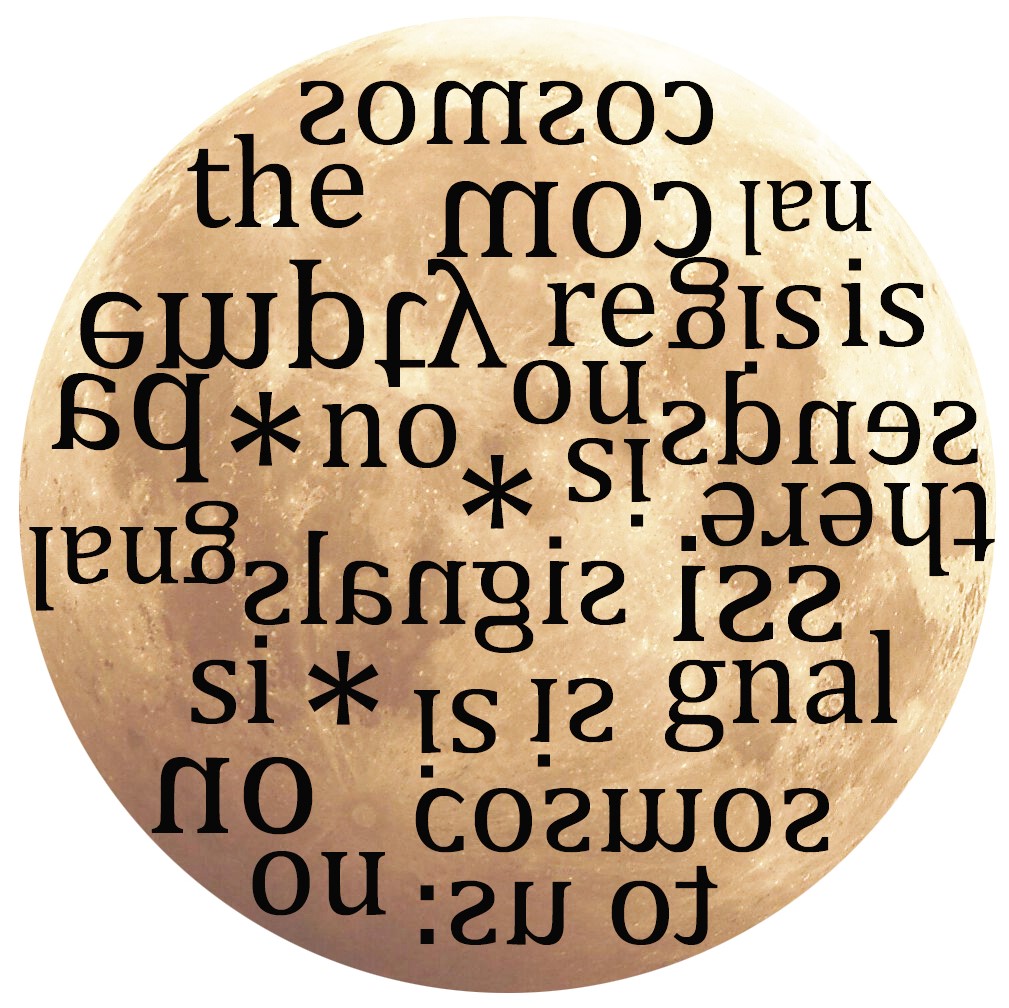}
\vspace{-0.2cm}
\caption{Illustration for the problem ``A space message''.}
\label{fig-planet}
\end{figure}

What message do you get (see Fig.\;\ref{fig-planet})?

\subsubsection{Solution}

It is easy to see fragments of words written in distinct directions and reflected vertically and horizontally. Once we combine all of them, we can read the message \begin{quote}
   {\sl cosmos sends signals to us:} \\
   {\sl compassion}\\
   {\sl there is no signal}\\
   {\sl cosmos is empty}\\
   {\sl there is no signal}\\
   {\sl no signal}
\end{quote}
So, between confusing messages, we can read what the cosmos is actually sending us: {\sl compassion}.

\subsection{Problem ``Two strings''}

\subsubsection{Formulation}
\hypertarget{pr-seq}{}

Carol takes inspiration from different strings and comes up with unusual ways to build them.
Today, she starts with a binary string $A_n$ constructed by induction in the following way.
Let $A_1 = 0$ and $A_2 = 1$. For $n > 2$, the string $A_n$ is defined by concatenating the strings $A_{n-1}$ and $A_{n-2}$ from left to right, i.\,e. $A_n = A_{n-1}A_{n-2}$.

Together with $A_n$ consisting of ``$0$''s and ``$1$''s, Carol constructs a ternary string $B_{n}$ consisting of ``$-1$''s, ``$0$''s and ``$1$''s. Let $A_n = a_1 ... a_m$ for appropriate $m$, where $a_i\in\{0,1\}$; then $B_n = b_1... b_{\ell}$, where $\ell = \lceil m/2 \rceil$ and $b_i\in\{-1,0,1\}$ is defined as follows:
$$b_i =  a_{2i-1} - a_{2i} \text{ for } i = 1,..., \ell \text{ (the exceptional case } b_{\ell} = a_m \text{ if } m \text{ is odd)}.$$

Help Carol to find all $n$ such that $B_n$ has the same number of ``$1$''s and ``$-1$''s.

\bigskip

\noindent{\bf Example.} The strings $A_n$ and $B_n$ for small $n$ are the following:
\begin{align*}
A_3 &= A_2A_1 = 10, & A_4 &= A_3A_2 = 101, &A_5 &= A_4A_3 = 10110, &A_6 &= A_5A_4 = 10110101. \\
B_3 &= 1,                 & B_4 &= 11,                &B_5 &= 100,  &B_6 &= 10(-1)(-1).   
\end{align*}

\subsubsection{Solution}

Let us consider $A_n$ as a decimal number. Then, by construction, the string $B_n$ has the same number of ``$1$''s and ``$-1$''s if and only if the number $A_n$ is divisible by 11. So, we need to find all $n$ such that $A_n = 0 \pmod{11}$.

The number of digits in $A_n$ is the $n$-th Fibonacci number $F_n$. It follows that $A_n$ modulo 11 satisfies a recursion:
$$A_n = 10^{F_{n-2}}A_{n-1} + A_{n-2} = (-1)^{F_{n-2}}A_{n-1} + A_{n-2} \pmod{11}.$$
It is easy to see that $F_n$ is even if and only if 3 divides $n$. Hence, $(-1)^{F_{n-2}}$ is periodic with period 3. Computing $A_n$ modulo 11 for small $n$ using the recursion, we find
$A_1,...,A_8 = 0, 1,-1, 2, 1, 1, 0, 1 \pmod{11}$.
By induction, we deduce that $A_{n+6} = A_n \pmod{11}$ for all $n$. Thus, $A_n$ is divisible by 11 if and only if $n = 1 \pmod{6}$.

\subsection{Problem ``Elliptic curve points''}

\subsubsection{Formulation}
\hypertarget{pr-residue}{}

Alice is studying elliptic curve cryptography. Her task for today is in practice with basic operations on elliptic curve points.
 Let $\mathbb F_p$ be the finite field with $p$ elements ($p>3$ prime). Let $E/\mathbb F_p$ be an elliptic curve in Weierstrass form, that is a curve with equation $y^2= x^3+ax+b$, where $a,b\in\mathbb F_p$ and $4a^3+27b^2\neq0$.
Recall that the affine points on $E$ and the point $\mathcal O$ at infinity form an abelian group, denoted
 $$
E(\mathbb F_p)=\{(x,y) \in \mathbb F_p^2\colon y^2=x^3+ax+b \} \cup\{\mathcal O\} \enspace.
$$

Assume that $b=0$. Let  $R\in E(\mathbb F_p)$ be an element of odd order, $R \neq \mathcal{O}$. 
Consider $H=\langle R\rangle$ that is the subgroup generated by $R$.

Help Alice to show that if $(u,v)\in H$, then $u$ is a quadratic residue mod $p$.

\begin{remark} For the Weierstrass form, $P_1 + P_2$ for $P_1, P_2 \in E(\mathbb{F}_p)$ is calculated as follows:
\begin{itemize}[noitemsep]
	\item $P_1 + \mathcal{O} = P_1$. 
	
	Next, we assume that $P_1, P_2 \neq \mathcal{O}$ and $P_1 = (x_1, y_1)$, $P_2 = (x_2, y_2)$.
	\item $P_1 + (-P_1) = \mathcal{O}$. Note that $-(x_1, y_1) = (x_1, -y_1)$. 
	
	Next, we assume that $P_1 \neq -P_2$.
	\item $P_1 + P_1 = P_3 = (x_3, y_3)$ can be calculated in the following way:
	$$
		x_3 = \frac{(3x_1^2 + a)^2}{(2y_1)^2} - 2x_1, \ \ \ \ \ \ \ \ \ y_3 = -y_1 - \frac{3x_1^2 + a}{2y_1} (x_3 - x_1).
	$$
	Next, we assume that $P_1 \neq P_2$.
	\item $ P_1 + P_2= P_3 = (x_3, y_3)$ can be calculated in the following way:
	$$
		x_3 = \frac{(y_2 - y_1)^2}{(x_2 - x_1)^2} - x_1 - x_2, \ \ \ \ \ y_3 = -y_1 - \frac{y_2 - y_1}{x_2 - x_1} (x_3 - x_1).
	$$
\end{itemize}
\end{remark}

\subsubsection{Solution}

Let $Q = (x', y') \in H \setminus \{\mathcal{O}\}$ and $n$ be the order of $Q$. It is clear that $n$ is a divisor of the order of $R$. Thus, $n$ is odd too. Next, it is straightforward that $Q = 2P$, where $P = (x, y) = \frac{n + 1}{2} Q$. Note that $y \neq 0$. Indeed, $(x,0) + (x,0) =  \mathcal{O}$, but $Q = 2(x,0) \neq \mathcal{O}$. By the formula of doubling points ($P + P$), we have that
	\begin{align*}
		x' &= \frac{(3x^2 + a)^2}{(2y)^2} - 2 x = \frac{9x^4 + 6ax^2 + a^2 - 8xy^2}{(2y)^2} = \frac{9x^4 + 6ax^2 + a^2 - 8x(x^3 + ax)}{(2y)^2}\\
		&= \frac{x^4 - 2ax^2 + a^2}{(2y)^2} = \Big(\frac{x^2 - a}{2y}\Big)^2.
	\end{align*}
	Hence, $x'$ is a quadratic residue mod $p$.

\subsection{Problem ``The number of rounds''}

\subsubsection{Formulation}
\hypertarget{pr-Nrounds}{}

A famous cryptographer often encrypts his personal data using his favourite block cipher.
The block cipher has three variants with $r_1 = 10$, $r_2 = 12$ and $r_3 = 14$ rounds. On this occasion, the cryptographer no longer remembers which of the variants he used.

Fortunately, the cryptographer did ask his students to write down the number of rounds for him.
However, in a creative mood, the students decided to encrypt it using a custom cipher $E_k$ with a 4-bit block size. As illustrated in Fig.~\ref{fig:studentScheme}, round $i$ of their construction XORs the $i$\textsuperscript{th} nibble $k_i$ of the key $k = k_1\|k_2\|\ldots\|k_{r + 1}$ with the state and then applies the function $S$ given in Table~\ref{tbl:studentSBox}. Lacking confidence in their own abilities, the students decided to instantiate the cipher $E_k$ with $r = r_1 \cdot r_2 \cdot r_3 + 1 = 1681$~rounds.

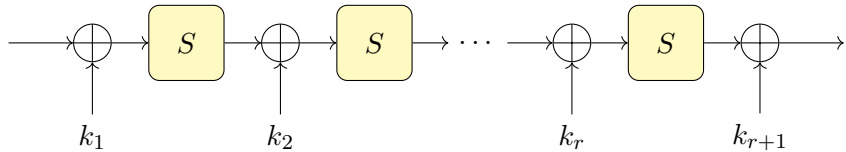
\begin{figure}[htb]
\centering
\begin{tikzpicture}[node distance=1.25cm]
\node (begin) [rectangle] {};
\node (k1)  [xor, right of = begin] {};
\node (r1) [box, right of = k1] {$S$};
\node (k2)  [xor, right of = r1] {};
\node (r2) [box, right of = k2] {$S$};
\node (dots) [rectangle, text centered, right of = r2, xshift=0.125cm] {\dots};
\node (k3)  [xor, right of = dots] {};
\node (r3) [box, right of = k3] {$S$};
\node (k4)  [xor, right of = r3] {};
\node (end) [rectangle, right of = k4] {};

\node (kt1)  [below of = k1] {$k_1$};
\node (kt2)  [below of = k2] {$k_2$};
\node (kt3)  [below of = k3] {$k_r$};
\node (kt4)  [below of = k4] {$k_{r+1}$};

\draw [->] (begin.east) -- (k1.west);
\draw [->] (k1.east) -- (r1.west);
\draw [->] (r1.east) -- (k2.west);
\draw [->] (k2.east) -- (r2.west);
\draw [->] (r2.east) -- (dots);
\draw [->] (dots) -- (k3.west);
\draw [->] (k3.east) -- (r3.west);
\draw [->] (r3.east) -- (k4.west);
\draw [->] (k4.east) -- (end);
\draw [->] (kt1) -- (k1.south);
\draw [->] (kt2) -- (k2.south);
\draw [->] (kt3) -- (k3.south);
\draw [->] (kt4) -- (k4.south);
\end{tikzpicture}
\label{fig:studentScheme}
\caption{The students' encryption method.}
\label{fig:studentScheme}
\end{figure}

\begin{table}[htb]
\centering\begin{tabular}{|c||cccccccccccccccc|}
\hline
$x$ &     \tt 0 & \tt 1 & \tt 2 & \tt 3 & \tt 4 & \tt 5 & \tt 6 & \tt 7 & \tt 8 & \tt 9 & \tt a & \tt b & \tt c & \tt d & \tt e & \tt f\\
\hline
$S(x)$ & \tt 3 & \tt e & \tt 6 & \tt 8 & \tt 0 & \tt c & \tt b & \tt 4 & \tt 1 & \tt d & \tt 5 & \tt a & \tt 7 & \tt 9 & \tt f & \tt 2\\
\hline
\end{tabular}
\caption{Lookup table for the function $S$ (in hexadecimal notation).} 
\label{tbl:studentSBox}
\end{table}

The students wrote down that the encryption of $r_1 = 10$ is $5$ and of $r_2 = 12$ is $0$, that is $E_k(1, 0, 1, 0) = (0, 1, 0, 1)$ and $E_k(1, 1, 0, 0) = (0, 0, 0, 0)$.
Of course, the students forgot the key, but they still remember that it was an ASCII-encoding of a passphrase consisting only of upper- and lower case English letters. After hearing this, the famous cryptographer exclaims that the students have made a mistake.

How did he know that something was wrong?

\subsubsection{Solution}
Let $u = (1, 1, 0, 0)$, then
\begin{equation*}
    u^\T S(x_1, x_2, x_3, x_4) = x_1 x_2 \oplus x_3 := f(x_1, x_2, x_3, x_4)\,.
\end{equation*}
In addition, it holds that
\begin{equation*}
    f(S(x_1, x_2, x_3, x_4) \oplus (0, 1, a, b)) = x_1 \oplus x_2 \oplus a \oplus 1 = u^\T{} x \oplus a \oplus 1\,.
\end{equation*}
Let $F_i = S(x \oplus k_i)$. Since the first four bits of an ASCII-encoding of an upper- or lowercase letter are always of the form $(0, 1, a, b)$, it follows that $k_i = (0, 1, a_i, b_i)$ for all odd $i$.
Hence, for odd $i$,
\begin{equation*}
    u^\T{} F_{i}(F_{i - 1}(x)) = f(F_{i - 1}(x) \oplus k_i) = u^\T (x \oplus k_{i - 1}) \oplus a_i \oplus 1\,,
\end{equation*}
Iterating this for an odd number of rounds $r$ shows that there exists a constant $c$ such that
\begin{equation*}
    u^\T(F_{r} \circ F_{r - 1} \circ \cdots \circ F_1)(x) = f(x) \oplus c\,.
\end{equation*}
The constant $c$ depends on the key.
Using the observations above, it is possible to show that the plaintext/ciphertext combinations reported by the students are incompatible:
\begin{enumerate}[noitemsep]
\item We have $f(1, 0, 1, 0) = 1$ and $u^\T(0, 1, 0, 1) = 1$. Hence, $c = 0$.
\item We have $f(1, 1, 0, 0) = 1$ and $u^\T(0, 0, 0, 0) = 0$. Hence, $c = 1$.
\item The above is a contradiction.
\end{enumerate}

\subsection{Problem ``Close to permutations''}

\subsubsection{Formulation}
\hypertarget{pr-xorplus}{}

Bob wants to use a new function inside the round transformation of a cipher.  He chooses a family $\mathcal{F}$ of functions $F_{\alpha}$ from $\mathbb{F}_2^n$ to itself of the form
 $$F_{\alpha}(x) = x \oplus (x \boxplus \alpha), \mbox{ where }$$
	\begin{itemize}[noitemsep]
		\item $x, \alpha \in \mathbb{F}_2^{n}$,
		\item $\oplus$ denotes the bit-wise XOR of binary vectors,
		\item $\boxplus$ denotes the addition modulo $2^n$ of integers whose binary representations are the given vectors.
	\end{itemize}	

Bob noted that functions from $\mathcal{F}$ are not bijective. So, he introduced a parameter that measures in some sense the closeness of a function to a permutation. For a given function $F$ from $\mathbb{F}_2^n$ to itself, the parameter is
	
$$C(F) = \#\{ (x, y) \in \mathbb{F}_2^n \times \mathbb{F}_2^n\ : \ F(x) = F(y) \}.$$

The smaller the parameter value, the  better the function. Bob wants to choose ``the best functions'' by this parameter among $\mathcal{F}$. Help Bob to find answers to the questions below!
	\begin{itemize}[noitemsep]
		\item[\bf Q1] How many ``the best functions'' exist in $\mathcal{F}$?
		\item[\bf Q2] What $\alpha$ correspond to ``the best functions'' from $\mathcal{F}$?
		\item[\bf Q3] What is $C(F_{\alpha})$ for ``the best functions'' from $\mathcal{F}$?
	\end{itemize}

\subsubsection{Solution}

First of all, we prove auxiliary results. Let $C(\alpha) = C(F_{\alpha})$ and $\alpha 0 = (0, \alpha_1, \ldots, \alpha_n)$ where $\alpha \in \mathbb{F}_2^n$. Similarly we define $\alpha1$. 
Before giving the answer, we prove that for any $\alpha$ the following properties hold:
\begin{align*}
	C(\alpha0) &= 4C(\alpha),\\
	C(\alpha1) &= C(\alpha) + C(\overline{\alpha}), \text{ where } \overline{\alpha} = (\alpha_1 \oplus 1, \ldots, \alpha_n \oplus 1),\\
	C(\alpha1) &< C(\alpha0).
\end{align*}

It is not difficult to see that $C(\alpha0) = 4C(\alpha)$, $\alpha \in \mathbb{F}_2^n$. Indeed, let us define $x' = (x_2, \ldots, x_{n+1}) \in \mathbb{F}_2^n$ for $x \in \mathbb{F}_2^{n+1}$. Since the first (the least significant) bits of both $x \oplus (x \boxplus  \beta)$ and $y \oplus (y \boxplus  \beta)$ are equal to $\beta_1$, where $x, y, \beta \in \mathbb{F}_2^{n+1}$, we can exclude them:
\begin{align*}
	C(\alpha 0) &= \#\{ x', y' \in \mathbb{F}_2^n, x_1, y_1 \in \mathbb{F}_2 : x' \oplus (x' \boxplus \alpha) = y' \oplus (y' \boxplus \alpha)\}
			    = 4 C(\alpha).
\end{align*}

Similarly we show that $C(\alpha1) = C(\alpha) + C(\alpha \boxplus 1)$. Indeed,
\begin{align}
	C(\alpha1) &= \#\{ x', y' \in \mathbb{F}_2^n, x_1 = 0, y_1 = 0 : x' \oplus (x' \boxplus \alpha) = y' \oplus (y' \boxplus \alpha)\} \label{eq:requrrence1}\\
			    &+ \#\{ x', y' \in \mathbb{F}_2^n, x_1 = 1 ,y_1 = 0 : x' \oplus (x' \boxplus \alpha \boxplus 1) = y' \oplus (y' \boxplus \alpha)\} \label{eq:requrrence2}\\
			    &+ \#\{ x', y' \in \mathbb{F}_2^n, x_1 = 0, y_1 = 1 : x' \oplus (x' \boxplus \alpha) = y' \oplus (y' \boxplus \alpha \boxplus 1)\} \label{eq:requrrence3}\\
			    &+ \#\{ x', y' \in \mathbb{F}_2^n, x_1 = 1, y_1 = 1 : x' \oplus (x' \boxplus \alpha \boxplus 1) = y' \oplus (y' \boxplus \alpha \boxplus 1)\}\label{eq:requrrence4}.
\end{align}
The first bit of $x' \oplus (x' \boxplus \alpha \boxplus 1)$ is equal to $\alpha_1 \oplus 1$, but the first bit of $y' \oplus (y' \boxplus \alpha)$ is equal to $\alpha_1$. It means that $(\ref{eq:requrrence2}) = 0$. Swapping $x'$ and $y'$, we obtain that $(\ref{eq:requrrence3}) = 0$ as well. Also, $(\ref{eq:requrrence1}) = C(\alpha)$ and $(\ref{eq:requrrence4}) = C(\alpha \boxplus 1)$.

Moveover, $C(\alpha1) = C(\alpha) + C(\overline{\alpha})$. The reasons are the following. It is clear that $\boxminus (\alpha \boxplus 1) = \boxminus \alpha \boxminus 1 = 2^n - 1 - \alpha = \overline{\alpha}$. Finally, for any $\beta \in \mathbb{F}_2^n$ it is true that
\begin{align*}
	C(\beta) &= \#\{ x, y \in \mathbb{F}_2^n : x \oplus (x \boxplus \beta) = y \oplus (y \boxplus \beta) \}\\
			   &= \#\{ x' = x \boxplus \beta, y' = y \boxplus \beta \in \mathbb{F}_2^n : (x' \boxminus \beta) \oplus x'  = (y' \boxminus \beta) \oplus y' \} = C(\boxminus \beta).
\end{align*}

Using these formulas, we show by induction that $C(\alpha) < 3 C(\overline{\alpha})$. The base of the induction $n = 1$ is straightforward since $C(0) = C(1) > 0$. Next, let $\alpha = \alpha'0$, $\alpha' \in \mathbb{F}_2^{n - 1}$. Then
$C(\alpha'0) = 4C(\alpha') = 3C(\alpha') + C(\alpha')$ and $3 C(\overline{\alpha'0}) = 3 C(\alpha') + 3 C(\overline{\alpha'})$. But $C(\alpha') <  3 C(\overline{\alpha'})$ by the induction hypothesis, which means that $C(\alpha'0) < 3C(\overline{\alpha'0})$. If $\alpha = \alpha'1$, then $C(\alpha'1) = C(\alpha') + C(\overline{\alpha'})$. At the same time, $3C(\overline{\alpha'1}) = 12 C(\overline{\alpha'}) = 11  C(\overline{\alpha'}) + C(\overline{\alpha'})$. The induction hypothesis shows that $C(\alpha'1) < C(\overline{\alpha'1})$. Finally, $C(\alpha) < 3 C(\overline{\alpha})$.

As a result, we can see that $C(\alpha1) = C(\alpha) + C(\overline{\alpha}) < C(\alpha) + 3C(\alpha) = C(\alpha0)$.

Now we can find minimum $\alpha^*_n$. First of all, it is straightforward that there is $2$ minimums for $n = 1,2$: $C(0) = C(1) = 4$ for $n = 1$ and $C(1) = C(3) = 8 < C(0) = C(2)$ for $n = 2$. Let $n > 2$.
We prove that there are $4$ minimums (Q1, $n > 2$), ``the best'' $\alpha^*_n$ from $\mathbb{F}_2^n$ is any of 
$$
	(1, \underbrace{\alpha_2, \overline{\alpha_2}, \alpha_2, \overline{\alpha_2}, \ldots }_{n - 2}, \alpha_n), \alpha_1, \alpha_n \in \mathbb{F}_2, \ (\text{Q2}, n > 2)
$$
and $C(\alpha^*_n) = C(\alpha^*_{n-1}) + 4 C(\alpha^*_{n - 2})$ (Q3, $n > 2$).

    Let us use induction by $n$. The base of the induction is mentioned above: any $\alpha \in \mathbb{F}_2$ provides $C(\alpha) = C(\alpha^*_1)$ and $C(\alpha1) = C(\alpha^*_2)$. Let us suppose that the answers are correct for $n$. Now we will prove that they hold for $n+1 \geq 3$. First of all, the first bit of $\alpha^*_{n + 1}$ is $1$ since $C(\alpha 1) < C(\alpha 0)$. Next, $C(\alpha 1) = C(\alpha ) + C(\overline{\alpha})$ for any $\alpha \in \mathbb{F}_2^n$. Let $\alpha' = (\alpha_2, \ldots, \alpha_n)$ if $\alpha_1 = 0$ and $\overline{\alpha'} = (\alpha_2, \ldots, \alpha_n)$ otherwise. Then
    \begin{align*}
            C(\alpha1) = C(\alpha'0) + C(\overline{\alpha'}1) = 4 C(\alpha') + C(\overline{\alpha'}1) \geq 4 C(\alpha^*_{n-1}) + C(\alpha^*_{n}).
    \end{align*}
It means that $C(\alpha 1) \geq C(\alpha^*_{n + 1}) \geq C(\alpha^*_{n}) + 4 C(\alpha^*_{n-1})$. Moreover, $C(\alpha 1) = C(\alpha^*_{n + 1}) = C(\alpha^*_{n}) + 4 C(\alpha^*_{n-1})$ if and only if $C(\alpha') = C(\alpha^*_{n-1})$ and $C(\overline{\alpha'}1) = C(\alpha^*_{n})$. By the induction hypothesis, these restrictions are equivalent to
    \begin{align}
            \alpha'_{i + 1} &= \overline{\alpha'_i} \text{ for any } i = 1, \ldots, n - 3, \text{ from } C(\overline{\alpha'}1) = C(\alpha^*_{n}), \label{eq:mininduction}\\
            \alpha'_{1} &= 1 \text{ for } n > 2, \text{ additionally from } C(\alpha') = C(\alpha^*_{n - 1}).\label{eq:mininduction2}
    \end{align} 
    Let $n > 2$. Since $\alpha'_1 = \alpha_2 \oplus \alpha_1$ by the definition of $\alpha'$, (\ref{eq:mininduction2}) guarantees that $\alpha_2 = \overline{\alpha_1}$. Taking into account~(\ref{eq:mininduction}), the restrictions transform to $\alpha_{i + 1} = \overline{\alpha_i}$ for $i = 1, \ldots, n - 2$. If $n = 2$, (\ref{eq:mininduction}) and (\ref{eq:mininduction2}) gives no restrictions. 
    Thus, all $\alpha 1$ with these restrictions and $\alpha^*_{n+1}$ coincide. The induction step is proven.

    There are famous techniques (see, for instance, \cite{Hor65}) to express the $n$-th element of a linear recurrence sequence. Thus, the answer $C(\alpha^*_{n}) = C(\alpha^*_{n - 1}) + 4 C(\alpha^*_{n-2})$ for (Q3) is equivalent to the following: 
    $$
        C(\alpha^*_{n})= \frac{1}{34 \cdot 2^n} \Big((17 + 7 \sqrt{17})(1 + \sqrt{17})^n + (17 - 7 \sqrt{17})(1 - \sqrt{17})^n\Big).
    $$    
	In addition, $C(\alpha)$ is connected with a special case of additive differential probabilities for the function $x \oplus y$, $x, y \in \mathbb{F}_2^n$, see~\cite{MouhaEtAl2021}.


\subsection{Problem ``A present for you!''}
This problem was given in two variants during the Olympiad. The original one that is described below was for ``university students'' and ``professionals''. A small variant of the problem was for ``school students''. It considered the bit permutation for 16 bits of the cipher {\sc small-present}.
\subsubsection{Formulation}
\hypertarget{pr-present}{}

\begin{figure}
\centering
\includegraphics[width=0.8\textwidth]{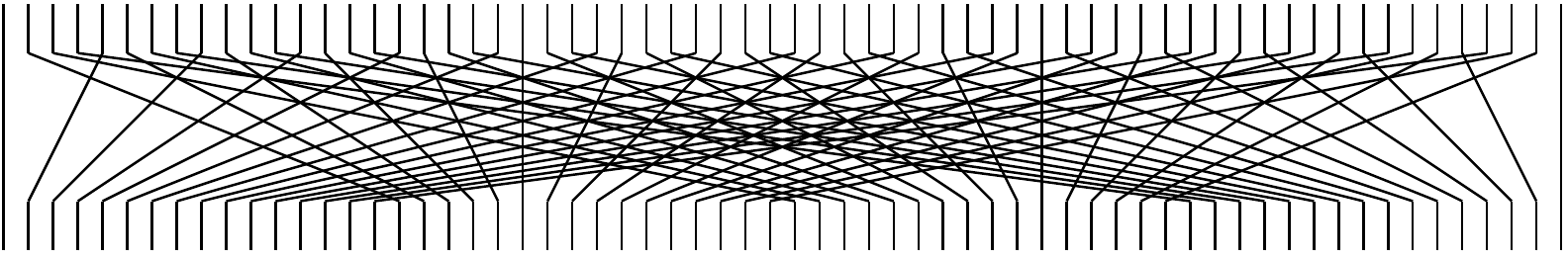}
\caption{Illustration of {\sc present} bit permutation.}
\label{fig:pLayer}
\end{figure}

Alice wants to implement the lightweight block cipher {\sc present} on a chip. She starts with the bit permutation that is defined in Table~\ref{tab:pLayer} and illustrated in Fig.~\ref{fig:pLayer}.
Clearly, many lines are intersecting, and this would cause a short circuit if the lines were metal wires. Is it possible to avoid this problem by using several ``layers,'' i.e., parallel planes? That is to draw the lines without intersections on each layer. We assume that
\begin{itemize}[noitemsep]
\item the work area is a rectangle bounded by the lines where input and output bits are placed and the lines of the outermost connections $P(0) = 0$ and $P(63) = 63$;
\item input and output bits are ordered; connections are represented by arbitrary curves;
\item color of a line indicates the number of its layer, a line can change color several times;
\item the point where a line changes color indicates a connection from one layer to another.
\end{itemize}

\begin{itemize}
\item[{\bf Q1}] What is the minimun number of layers required for implementing in this way the {\sc present} bit permutation?
\item[{\bf Q2}] Find a systematic approach how to draw a valid solution for the minimum number of layers found in {\bf Q1} and present the drawing!

For your help (but not necessarily), you can use  \href{https://app.diagrams.net/}{\textcolor{blue}{a specific online tool}} \cite{diagrams} and \href{https://nsucrypto.nsu.ru/media/MediaFile/present-orig.drawio}{\textcolor{blue}{download}} \cite{nsucrypto-present} the {\sc present}  bit permutation as in Figure.
\end{itemize}

\begin{table}[h]
\centering
\scalebox{0.9}{
\begin{tabular}{|c||c|c|c|c|c|c|c|c|c|c|c|c|c|c|c|c|}
\hline
$i$    &  0 &  1 &  2 &  3 &  4 &  5 &  6 &  7 &  8 &  9 & 10 & 11 & 12 & 13 & 14 & 15\tabularnewline
$P(i)$ &  0 & 16 & 32 & 48 &  1 & 17 & 33 & 49 &  2 & 18 & 34 & 50 &  3 & 19 & 35 & 51\tabularnewline
\hline
\hline
$i$    & 16 & 17 & 18 & 19 & 20 & 21 & 22 & 23 & 24 & 25 & 26 & 27 & 28 & 29 & 30 & 31\tabularnewline
$P(i)$ &  4 & 20 & 36 & 52 &  5 & 21 & 37 & 53 &  6 & 22 & 38 & 54 &  7 & 23 & 39 & 55\tabularnewline
\hline
\hline
$i$    & 32 & 33 & 34 & 35 & 36 & 37 & 38 & 39 & 40 & 41 & 42 & 43 & 44 & 45 & 46 & 47\tabularnewline
$P(i)$ &  8 & 24 & 40 & 56 &  9 & 25 & 41 & 57 & 10 & 26 & 42 & 58 & 11 & 27 & 43 & 59\tabularnewline
\hline
\hline
$i$    & 48 & 49 & 50 & 51 & 52 & 53 & 54 & 55 & 56 & 57 & 58 & 59 & 60 & 61 & 62 & 63\tabularnewline
$P(i)$ & 12 & 28 & 44 & 60 & 13 & 29 & 45 & 61 & 14 & 30 & 46 & 62 & 15 & 31 & 47 & 63\tabularnewline
\hline
\end{tabular}
}
\caption{Definition of the bit permutation used in {\sc present}. Bit $i$ is moved to bit position $P(i)$.}
\label{tab:pLayer}
\end{table}

\subsubsection{Solution}

A first attempt to solve this problem, would be to try and connect some inputs and outputs. However, it will not take long to get stuck without a systematic approach.

\begin{figure}
\centering
\begin{tabular}{cc}
\includegraphics[scale=0.58]{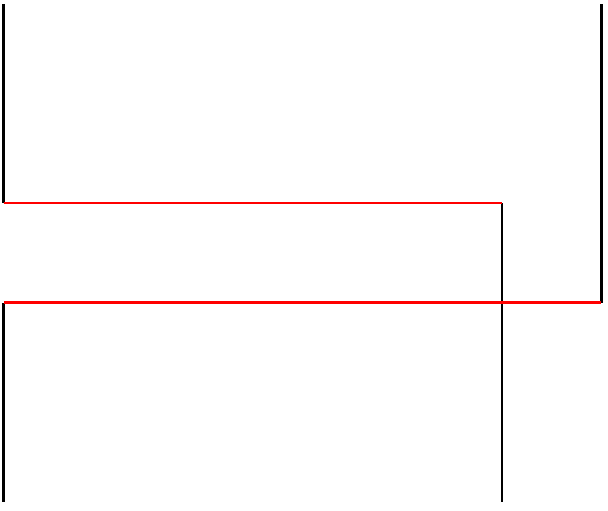}
&
\includegraphics[scale=0.75]{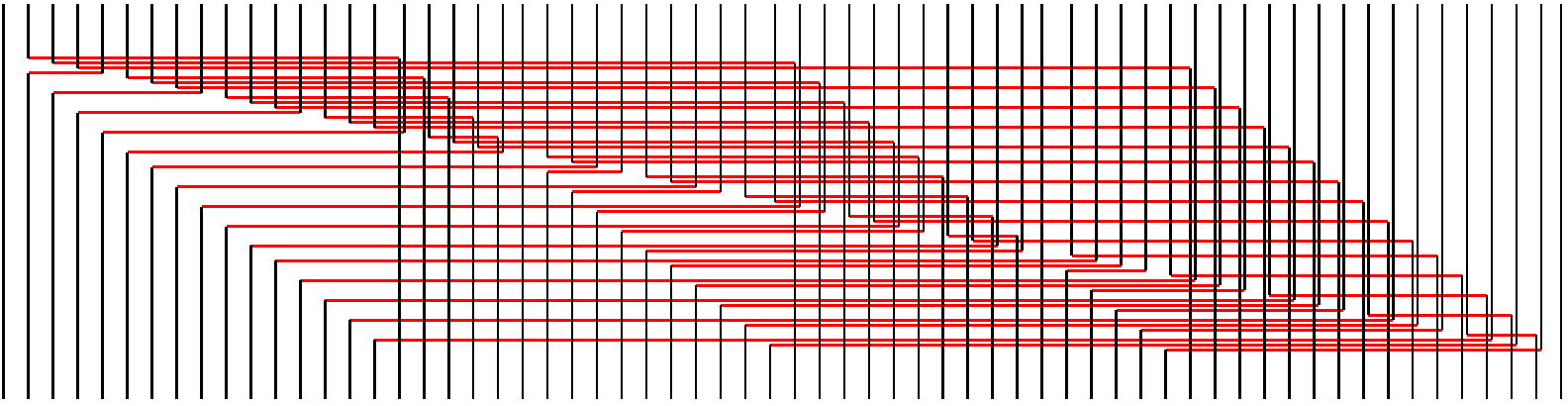}
\\
(a) Swapping two lines.
&
(b) Illustration of {\sc present} bit permutation using two layers.
\end{tabular}
\caption{Illustrations to the solution of the problem ``A present for you!''}
\label{fig:present}
\end{figure}

An observation is that lines with several different angles create a problem, as it becomes difficult to predict where they might intersect with other lines. A way to overcome this is to work with only horizontal and vertical lines. The vertical lines can be in one color, and the horizontal lines in another color. This approach gives us an idea to use two layers. Let us show how to draw a scheme.
All lines of the same color are parallel, however some lines might overlap. To see how to address this, consider the simple case of swapping two inputs, as shown in Fig.~\ref{fig:present}\,(a). As the drawing shows, overlapping lines can be avoided by moving the second input slightly to the right. This is just done to make the drawing a bit easier; note that it does not affect the validity of the solution as the order of the inputs is preserved.

This method can be extended to an arbitrary number of inputs. A full solution for the {\sc present} bit permutation is given in Fig.~\ref{fig:present}\,(b).

\subsection{Problem ``Nonlinear hiding''}

\subsubsection{Formulation}
\hypertarget{pr-hiding}{}

Nicole is learning about secret sharing. She created a binary vector $y \in \F_2^{\W}$ and splitted it into $\K$ shares $x_i \in \F_2^{\W}$ (here $\oplus$ denotes the bit-wise XOR):
$$
	y = x_1 \oplus x_2 \oplus ... \oplus x_{\K}.
$$
Then, she created $\K$ more random vectors $x_{\Kplus},...,x_{\Ktwo}$ and shuffled them together with the shares $x_1, ...,x_{\K}$. Formally, she chose a secret permutation $\sigma$ of $\{1,...,\Ktwo\}$ and computed
\begin{align*}
	z_1 &= x_{\sigma(1)},\\
	z_2 &= x_{\sigma(2)},\\
	&...\\
	z_{\Ktwo} &= x_{\sigma(\Ktwo)},\\
\end{align*}

\vspace{-0.5cm}
\noindent where each vector $z_i \in \F_2^{\W}$. Finally, she splitted each $z_i$ into 5-bit blocks, and applied a secret bijective mapping $\rho: \F_2^5 \to \mathcal{S}$, where
$$
\mathcal{S} = \{\texttt{0},\texttt{1},\texttt{2},\texttt{3},\texttt{4},\texttt{5},\texttt{6},\texttt{7},\texttt{8},\texttt{9},\texttt{a},\texttt{b},\texttt{c},\texttt{d},\texttt{e},\texttt{f},\texttt{g},\texttt{h},\texttt{i},\texttt{j},\texttt{k},\texttt{l},\texttt{m},\texttt{n},\texttt{o},\texttt{p},\texttt{q},\texttt{r},\texttt{s},\texttt{t},\texttt{u},\texttt{y}\}
$$
(this strange alphabet has \texttt{y} instead of \texttt{v}).

Formally, she computed $Z_i \in \mathcal{S}^{\w}, 1 \leqslant i \leqslant \Ktwo$ such that
$$
	Z_i = (\rho(z_{i,1\ldots5}), \rho(z_{i,6\ldots10}), ..., \rho(z_{i,6556\ldots6560})).
$$

After Nicole came back from school, she forgot all the details! She only has written all the~$Z_i$ and she also remembers the first 6432 bits of $y$ (128 more are missing).
The \href{https://nsucrypto.nsu.ru/media/MediaFile/data-sharing.txt}{\textcolor{blue}{attachment}} \cite{nsucrypto-file-hiding} contains the 6432-bit prefix of $y$ on the first line and $Z_1,..., Z_{40} \in \mathcal{S}^{1312}$ on the following lines, one per line.

Help Nicole to recover full $y$!

\subsubsection{Solution}


This problem is inspired by the setting of generic white-box attacks \cite{18-Udovenko}. Consider an obfuscated program, where a secret function is protected by a linear masking scheme (secret sharing), and the shares are scattered among fully random values. In addition, each value is protected by a fixed random S-box (so called \emph{encoding}). The goal of an adversary is to recover the full secret function from a partial knowledge of it on a few inputs, just by observing all the described values.

In the Olympiad's problem, each row $Z_i$ corresponds to a chosen share or a random value, and each column corresponds to a distinct ``execution'' (i.e., a recording of values on a distinct input of the program). 

This problem can be solved by formulating the problem as a quadratic system of equations over $\mathbb{F}_2$ and solving it through linearization. More precisely, introduce 40 variables $t_i \in \mathbb{F}_2$, one per each row $i, 1 \leqslant i \leqslant 40$, describing whether the $i$-th row is a secret share. In addition, introduce 32 variables $m_c \in \mathbb{F}_2$, one per each $c \in \mathcal{S}$, describing the first bit of $\rho^{-1}(c)$. 
Then, each known 5-bit chunk $y_{5j+1\ldots 5j+5}$ of $y$ (more precisely, its first bit) gives a quadratic equation \[
\text{equation}~j, 1 \leqslant j \leqslant 1286:~~~~\bigoplus_{1\leqslant i \leqslant 40} t_i\cdot m_{Z_{i,j}} = y_{5j+1}.
\]
This system can be linearized. More precisely, introduce a new variable $w_{i,j} = t_i \cdot m_{Z_{i,j}} \in \F_2$ per each monomial $t_i \cdot m_{Z_{i,j}}$. There are $40 \times 32 = 1280$ variables and $6432 / 5 \geqslant 1286$ equations. After solving this linear system, we can see which rows $Z_i$ correspond to the shares of $y$ and a mapping defining first coordinate of $\rho^{-1}$ (up to a constant), allowing to recover every 5-th bit of the missing part. Repeating this procedure for 4 other positions allows to fully recover the value (note that the values of $t_i$ would already be recovered). 

Also, there was a hidden text in the random beginning prefix of $y$ dedicated to the 100th anniversary of the Cryptographic Service of Russian Federation:
\begin{quote}
   {\sl 2021 marks the centenary of the cryptographic service in Russia! On May 5, 1921, the 8th special department was created. Its tasks included the study of theoretical problems of cryptography and the development of new ciphers, the organization of cipher communication, cryptanalysis, radio monitoring and radio interception, etc. }
\end{quote}

\subsection{Problem ``Let's decode!''}

\subsubsection{Formulation}
\hypertarget{pr-decode}{}

Bob realized a cipher machine for encoding integers from  $0$ to $n-1$ by 128-bit strings using the secret function $\texttt{Enc}$. He set $n=1060105447831$. The cipher machine works as follows: it takes as input a pair of non-negative decimal integers $x$ and $d$ and returns  
$$
\texttt{Enc}(x^d\bmod n).
$$

Bob chose a secret number $k$ from $0$ to $n-1$ and asked Alice to guess it. Alice said that she can find~$k$ if Bob provides her with the cipher machine with an additional property. Namely, $x$ can be also of the form ``$k$'', and then the cipher machine will return
$\texttt{Enc}(k^d \bmod n)$.
In particular, for the query ``$k, 1$'', the cipher machine returns 
$$
\texttt{Enc}(k)=\texttt{41b66519cf4356cbbb4e88a4336024da}
$$
(the result is in hexadecimal notation). The cipher machine is \href{https://nsucrypto.nsu.ru/olymp/2021/round/2/task/4}{\textcolor{blue}{here}} \cite{nsucrypto-cipher}.

Prove Alice is right and find $k$ with as few requests to the cipher machine as possible!

\subsubsection{Solution}

We present three ways to solve the problem.

\smallskip
\noindent {\bf The first way (the authors' one).}
For the request ``$0, 1$'', we get the answer that is not equal to $\texttt{Enc}(k)$. Hence, $k\neq 0$.
A nonzero key $k$ can be represented as $g^x\bmod n$, where $g=12$ is 
a primitive root modulo~$n$. It remains to determine $x\in\{0,1,...,n-1\}$.

To find $x$, we apply the Pohlig\,---\,Hellman method. 
We use the fact that
$$
n-1=2\cdot 3\cdot 5\cdot 11\cdot 13\cdot 17\cdot 19\cdot 23\cdot 29\cdot 
31\cdot 37
$$
is a smooth number (all its prime factors are small).
The method works as follows:
\begin{enumerate}[noitemsep]
\item
For each prime factor $p$ of $n-1$, we find $x_p = x \bmod p$. 

To do this, we calculate $\texttt{Enc}(k^{(n-1)/p}\bmod n)$
and then $\texttt{Enc}(g^{i(n-1)/p}\bmod n)$, $i=0,1,\ldots,p-1$. 
The equality
$$
\texttt{Enc}(g^{i(n-1)/p}\bmod n)=\texttt{Enc}(k^{(n-1)/p}\bmod n)
$$
means that $x_p=i$.

The number of requests to $\texttt{Enc}$ can be reduced if 
the baby-step giant-step method is applied.

\item Obtaining all $x_p$, we solve the Chinese remainder problem 
$\{x\equiv x_p\bmod p\colon p \mid (n-1)\}$.
\end{enumerate}

The answer is $k = 856182870494$. 

\smallskip
\noindent {\bf The second way.} 
This solution is based on extracting roots modulo $n$. 
Let $k_0=1$, 
$$
k_i = k^{(n-1)/(p_1 p_2\ldots p_i)}\bmod n,\ i=1,2,\ldots,r,
$$
where $r=11$ is the number of prime factors of $n-1$ and
$p_1,p_2,\ldots,p_r$ are these factors.

The number $k_i$ is the $p_i$-th root of $k_{i-1}$.
These roots can be found by factoring the polynomial 
$x^{p_i}-k_{i-1}$ over the finite field of order $n$.
The required root $k_i$ is the one satisfying 
$$
\texttt{Enc}(k_i^1\bmod n)=\texttt{Enc}(k^{(n-1)/(p_1 p_2 \ldots p_i)}\bmod n).
$$
The key $k$ is the last root $k_r$.


\smallskip
\noindent {\bf The third way.} 
Let $k_i=k^{(n-1)/p_i}\bmod n$, $i=1,2,\ldots,r$. 
The number $k_i$ is the $p_i$-th root of $1$. 
It can be determined by comparing the codes of all possible 
roots with the code $\texttt{Enc}(k^{(n-1)/p_i}\bmod n)$. 

After determining the numbers $k_i$, we solve the system
$$
k^{a_i}\equiv k_i\pmod{n},\quad
a_i=(n-1)/p_i.
$$
To do this, we use Bezout's identity
$$
\sum_{i=1}^r a_i b_i = 1.
$$
Here $b_i$ are integer coefficients that can be determined using the extended 
Euclidian algorithm. 
Finally,
$$
k = k^{\sum_i a_i b_i} \bmod n = \prod_i k_i^{b_i}\bmod n.
$$


\subsection{Problem ``Shuffle ballots''}

\subsubsection{Formulation}
\hypertarget{pr-ballots}{}

In electronic voting, $n$ voters take part. Each of them is assigned a {\bf unique identifier} that is a number from the set $\{0,1,\ldots,n-1\}$. 
Shuffling of ballots during elections is implemented through the encryption of identifiers. When encrypting, the following conditions must hold:

\begin{enumerate}[noitemsep]
\item[{\bf 1.}]
The encryption result is again an integer from $\{0,1,\ldots,n-1\}$.
\item[{\bf 2.}]
The encryption process must involve the block cipher AES with a fixed key~$K$.  
\item[{\bf 3.}]
The number of requests to $\text{AES}_K$ must be the same for each identifier. 
\item[{\bf 4.}]
In order to manage security assurances, it should be possible to customize the number of requests to $\text{AES}_K$.
\end{enumerate}

Suggest a way how to organize the required encryption process of identifiers for
 $n=5818342$  and $n=5818343$. In other words, propose a method for organizing a bijective mapping from $\{0,1,\ldots,n-1\}$ to itself that satisfies conditions described above.

\subsubsection{Solution}

{\bf Case 1.}
The number $n=5818342$ is composite. It is factored as a product of 
numbers close to each other, namely $n_1=2594$ and $n_2=2243$. 
Hence, an identifier $x\in\{0,1,\ldots,n-1\}$ can be uniquely represented as 
$x = x_1 n_2 + x_2$, where $x_1\in\{0,1,\ldots,n_1-1\}$ and $x_2\in\{0,1,\ldots,n_2-1\}$. 

We can encrypt identifiers by applying several rounds of the form:
$$
(x_1, x_2)\leftarrow\bigl(y_1,(x_2 + \text{AES}_K(y_1+\beta))\bmod n_2\bigr),
\quad
y_1=(x_1 + \text{AES}_K(x_2+\alpha))\bmod n_1.
$$
Here $\alpha$, $\beta$ are round constants. 
We process numbers with $\text{AES}_K$ encoding them in 128-bit blocks before 
encrypton and decoding back after.

The proposed construction follows the UNF (Unbalanced Number Feistel) scheme 
\cite{ballots}.
When $n_1\approx n_2$ (that is our case), at least 3 rounds should be used 
to ensure security. Generally speaking, security guarantees are strengthened 
with increasing the number of rounds. 

{\bf Case 2.} 
The number $n=5818343$ is prime. So, the UNF scheme cannot be directly applied. 
Nevertheless, we can reduce the problem to the UNF encryption for a composite 
modulus $n'= n-1$ that was considered in Case 1 above.
We act as follows:
\begin{enumerate}[noitemsep]
\item
A number $a$ is chosen at random from the set $\{0,1,\ldots,n-1\}$.
\item
Suppose we need to encrypt $x \in \{0,1,\ldots,n-1\}$. 
If $x \neq a$, then we determine
$$
x' = \begin{cases}
x, & x < a;\\
x - 1, & x > a. 
\end{cases}
$$
The number $x'$ belongs to the set $\{0,1,\ldots,n'-1\}$. 
We encrypt $x'$ using the UNF scheme with $d$ rounds.
\item
If $x = a$, then we assign to $x$ the ciphertext $n'=n-1$. 
Additionally, to satisfy Requirement~3 for a constant number of requests to AES,  
we perform $d$ dummy AES encryptions.
Note that Requirement 3 is a countermeasure against timing attacks.
\end{enumerate}

We would like to briefly present ideas proposed by the participants.

\bigskip
\noindent{\bf The first idea.}
The prime $n$ is incremented rather than decremented. 
Using UNF, we construct a bijection $E_K$ on $\{0,1,\ldots,n\}$.
Then we encrypt $x\neq n$ with $E_K$ and get $y\neq n$.
What should we do if $E_K(x)=n$? There are 3 possiblities:
\begin{enumerate}[noitemsep]
\item
Precalculate $x_0=E_K^{-1}(n)$.
If $x=x_0$, then return $E_K(n)$.
If $x\neq x_0$, then return $E_K(x)$.

\item
Precalculate $y_0=E_K(n)$.
If $y=E_K(x)$ is equal to $n$, then return $y_0$.
Otherwise, return $y$.

\item
Without precalculations.
Calculate $y=E_K(x)$ and $z=E_K(y)$.
If $y=n$, then return $z$.
Otherwise, return $y$.
\end{enumerate}

\smallskip
\noindent{\bf The second idea.}
The encryption can be given by a permutation polynomial over the integer ring modulo $n$.
For example,
$$
f_K(x)=(\ldots((x+k_1)^e + k_2)^e +\ldots+k_{r-1})^e + k_r)\bmod n.
$$
Here $k_1, k_2,\ldots, k_r$ are round keys which are built using $\text{AES}_K$ 
(for instance, $k_i=\text{AES}_K(i)\bmod n$) and $e$ is coprime with 
$\varphi(n)$. 
We are dealing with the composition of permutations $x\mapsto x^e\bmod 
n$ and $x\mapsto (x + 1)\bmod n$ which is itself a permutation.

\subsection{Problem ``Studying Feistel schemes''}

\subsubsection{Formulation}
\hypertarget{pr-feistel}{}

The classical Feistel scheme and its generalizations are widely used to construct iterated block ciphers. {\bf Generalized Feistel schemes} (GFS) usually divide a message into $m$  subblocks and applies the (classical) Feistel transformation for a fixed number of two subblocks, and then performs a cyclic shift of $m$  subblocks.

Trudy wants to compare algebraic properties of different generalizations of the Feistel scheme based on shift registers over an arbitrary finite commutative ring with identity. For studying, she chooses a nonlinear feedback shift register (NLFSR), Type-II GFS and Target-Heavy (TH) GFS. She wants to decide whether or not these transformations belong to the {\bf alternating group} (that is the group of all even permutations). Trudy needs your help!

Let us give necessary notions. By $A(X)$ we denote the alternating group on a set $X$.
Let $t$ be a positive integer, $t \geqslant 1$, $(R, + , \cdot )$ be a commutative ring with identity 1, $\left| R \right| = {2^t}$.  The characteristic ${\rm{char(}}R{\rm{)}}$ of $R$ is equal to ${2^c}$ for some $c \in \{ 1,...,t\} $. In many block ciphers, we have
$$R \in \left\{{\mathbb{Z}_2^t,\ \mathbb{Z}_{2^t},\ \textbf{GF}(2^t)} \right\}, \; {\rm{char}}{(\mathbb{Z}_{2^t})} = 2^t, \; {\rm{char}}\left( {{\mathbb{Z}_{2}^t}} \right) = {\rm{char}}\left({\textbf{GF}(2^t)} \right) = 2.$$

\begin{enumerate}
  \item[{\bf Q1}]{\textbf{NLFSR}. Let $\ell \geqslant 1$, $m = {2^\ell}$, $h:{R^{m - 1}} \to R$. Consider a mapping $g_{k,h}^{({\rm{NLSFR}})}:{R^m} \to {R^m}$ defined by
$$g_{k,h}^{({\rm{NLSFR}})}:({\alpha _1},...,{\alpha _m}) \mapsto \left( {{\alpha _2},{\alpha _3},...,} \right.\left. {{\alpha _{m - 1}},{\alpha _m},{\alpha _1} + h({\alpha _2},...,{\alpha _m}) + k} \right)$$
for all $({\alpha _1},...,{\alpha _m}) \in {R^m}$, $k \in R$.  Describe all positive integers $t \geqslant 1$, $\ell,c \geqslant 1$ and a mapping $h:{R^{m - 1}} \to R$ such that $g_{k,h}^{({\rm{NLSFR}})} \in A({R^m})$  for any $k \in R$. Prove your answer!
  }
  \item[{\bf Q2}] {\textbf{Type-II GFS}. Let $\ell \geqslant 2$, $m = {2^\ell}$, $h = ({h_1},...,{h_{m/2}})$, where ${h_i}:R \to R$ for  $1 \leqslant i \leqslant m/2 $. Consider a mapping $g_{k,h}^{({\rm{GFS - II}})}:{R^m} \to {R^m}$ defined by
\begin{multline*}
g_{k,h}^{({\rm{GFS - II}})}:({\alpha _1},...,{\alpha _m}) \mapsto \ 
( {\alpha _2} + {h_1}({\alpha _1}) + {k_1},{\alpha _3},{\alpha _4} + {h_2}({\alpha _3}) + {k_2},\alpha _5,..., \\
 {\alpha _{m - 1}},{\alpha _m} + {h_{m/2}}({\alpha _{m - 1}}) + {k_{m/2}},{\alpha _1} )
\end{multline*}
for all $({\alpha _1},...,{\alpha _m}) \in {R^m}$, $k = ({k_1},...,{k_{m/2}}) \in {R^{m/2}}$. Describe all positive integers $t \geqslant 2$, $\ell,c \geqslant 1$ and mappings ${h_1},...,{h_{m/2}}$  such that $g_{k,h}^{({\rm{GFS - II}})}~\in~A({R^m})$  for any $k \in {R^{m/2}}$. Prove your answer!
   }
  \item[{\bf Q3}] {\textbf{TH-GFS}. Let $\ell \geqslant 2$, $m = {2^\ell}$, $h = ({h_2},...,{h_m})$, where  ${h_i}:R \to R$ for $2 \leqslant i \leqslant m$. Consider a mapping $g_{k,h}^{({\rm{TH}})}:{R^m} \to {R^m}$ defined by
\begin{multline*}
g_{k,h}^{({\rm{TH}})}:({\alpha _1},...,{\alpha _m}) \mapsto \left( {{\alpha _2} + {h_2}({\alpha _1}) + {k_2},{\alpha _3} + {h_3}({\alpha _1}) + {k_3},...,} \right.\\
\left. {{\alpha _{m - 1}} + {h_{m - 1}}({\alpha _1}) + {k_{m - 1}},{\alpha _m} + {h_m}({\alpha _1}) + {k_m},{\alpha _1}} \right)
\end{multline*}
for all $k = ({k_2},...,{k_m}) \in {R^{m - 1}}$. Describe all positive integers $t \geqslant 2$, $\ell,c \geqslant 1$ and mappings ${h_2},...,{h_m}$ such that $g_{k,h}^{({\rm{TH}})} \in A({R^m})$  for any $k \in {R^{m - 1}}$. Prove your answer!
   }
\end{enumerate}

\subsubsection{Solution}

Let $s$ be a permutation on a set $X$ with $v$ disjoint cycles of lengths ${\ell_1},...,{\ell_v}$.  By $\tau (s)$ denote
$$\tau (s) = {\ell_1} + ... + {\ell_v} - v.$$
Let ${\rm{sign}}(s)$ denote the sign of $s$. It is well known that ${\rm{sign}}(s) = {( - 1)^{\tau (s)}}$, i.e. $s$ is even and belongs to the alternating group $A(X)$ on $X$ if $\tau (s) \equiv 0\;(\bmod \;2)$. Moreover, ${\rm{sign}}: S(X) \to \{  - 1,1\} $ is a homomorphism, i.\,e. for all $s,b \in S(X)$, it holds
$${\rm{sign}}(sb) = {\rm{sign}}(s) \cdot {\rm{sign}}(b).$$

Let ${\rm{ord}}(b)$ be the order of $b$ in additive group $(R, + )$.

\noindent \textbf{Q1 NLSFR}.
Consider three permutations $\rho :{R^m} \to {R^m}$, $\delta _1^{(i)}:{R^m} \to {R^m}$, ${\theta _h}:{R^m} \to {R^m}$ defined for all $({\alpha _1},...,{\alpha _m}) \in {R^m}$ by the following rules:
\begin{align*}
\delta _1^{(i)}&:({\alpha _1},...,{\alpha _m}) \mapsto ({\alpha _1},...,{\alpha _{i - 1}},{\alpha _i} + 1,{\alpha _{i + 1}},...,{\alpha _m}), \; i \in \{ 1,...,m\},\\
\rho &:({\alpha _1},...,{\alpha _m}) \mapsto ({\alpha _2},...,{\alpha _m},{\alpha _1}),\\
{\theta _h}&:({\alpha _1},...,{\alpha _m}) \mapsto ({\alpha _1} + h({\alpha _2},...,{\alpha _m}),{\alpha _2},...,{\alpha _m})
\end{align*}

It is clear that
$$\delta _k^{(i)} = {\left( {\delta _1^{(i)}} \right)^k}~\text{ and }~g_{k,h}^{({\rm{NLSFR}})} = \rho \delta _k^{(1)}{\theta _h}.$$
To find ${\rm{sign}}\left( {g_{k,h}^{({\rm{NLSFR}})}} \right)$, we compute ${\rm{sign}}\left( {{\theta _h}} \right)$, ${\rm{sign}}\left( {\delta _1^{(i)}} \right)$, ${\rm{sign}}(\rho )$ and finally $\tau \left( {{\theta _h}} \right)$. 

Let us find $\tau \left( {{\theta _h}} \right)$. By definition, put
$${r_i} = \left| {\left\{ {({\beta _1},...,{\beta _{m - 1}}) \in {R^{m - 1}}\ |\ {\rm{ord}}\left( {h({\beta _1},...,{\beta _{m - 1}})} \right) = {2^i}} \right\}} \right|$$
for all $i \in \{ 0,...,c\} $. It is obvious that
$$\theta _h^j:({\alpha _1},...,{\alpha _m}) \mapsto \left( {{\alpha _1} + j \cdot h({\alpha _2},...,{\alpha _m}),{\alpha _2},...,{\alpha _m}} \right)$$
for any $({\alpha _1},...,{\alpha _m}) \in {R^m}$.  
Hence, the length of a cycle of ${\theta _h}$ is equal to ${2^i}$ for some $i \in \{ 0,...,c\} $. The number of cycles of length ${2^i}$ is $\left| R \right|{r_i} = {2^{t - i}}{r_i}$. Therefore,
$$\tau \left( {{\theta _h}} \right) = \sum\limits_{j = 1}^c {{2^{t - j}}{r_j}({2^j} - 1)} .$$
Thus,
$$
{\rm{sign}}\left( {{\theta _h}} \right) =
\begin{cases}
    -1, &\text{if } c = t, \; {r_c} \equiv 1\,(\bmod \,2), \\
    1,  &\text{if }  c<t,\; c = t,\; r_c \equiv 0\,(\bmod \,2).
\end{cases}
$$

Now we find $\tau (\delta _1^{(i)})$ for $0 \leqslant i \leqslant m$. Note that we have
$${\left( {\delta _1^{(i)}} \right)^k}:({\alpha _1},...,{\alpha _m}) \mapsto ({\alpha _1},{\alpha _2},...,{\alpha _{i - 1}},{\alpha _i} + k,{\alpha _{i + 1}}...,{\alpha _m})$$
for all $({\alpha _1},...,{\alpha _m}) \in {R^m}$, $k \in R$. So, $\delta _1^{(i)}$ has ${2^{c(m - 1)}}$ cycles of length ${2^c}$. Therefore, $\tau (\delta _1^{(i)}) = {2^{mt}} - {2^{c(m - 1)}}$, i.\,e. ${\rm{sign}}\left( {\delta _1^{(i)}} \right) = 1$.

It is well known \cite{Zie97} that $\rho $ has ${2^t}^{{2^j} - j} - {2^{t{2^{j - 1}} - j}}$ cycles of length ${2^j}$ for $0 \leqslant j \leqslant \ell$. Thus,
$$\tau (\rho ) = \sum\limits_{j = 1}^l {{2^{t{2^{j - 1}} - j}}({2^j} - 1)\left( {{2^t}^{{2^{j - 1}}} - 1} \right)} .$$

Hence,
$$
{\rm{sign}}\left( {\rho} \right) =
\begin{cases}
    -1, &\text{if } \ell=t=1, \\
    1, &\text{if } \ell \geqslant 2 \text{ or }  \ell=1,\;t \geqslant 2.
\end{cases}
$$

Therefore,
$${\rm{sign}}\left( {g_{k,h}^{({\rm{NLSFR}})}} \right) = {\rm{sign}}\left( \rho  \right) \cdot {\rm{sign}}\left( {\delta _k^{(1)}} \right) \cdot {\rm{sign}}\left( {{\theta _h}} \right) 
 = {\rm{sign}}\left( \rho  \right) \cdot {\rm{sign}}\left( {{\theta _h}} \right) = $$
$$=
\begin{cases}
    -1, &\text{if $\ell=t=c=1,\; r_c \equiv 0\,(\bmod \,2), $} \\
    -1, &\text{if $t=c,\; r_c \equiv 1\,(\bmod \,2),\; \ell\cdot t\geqslant 2, $} \\
     1, &\text{if $c<t$ or $\ell=t=c=1,\; r_c \equiv 1\,(\bmod \,2),$} \\
     1, &\text{if $t=c,\; r_c \equiv 0\,(\bmod \,2),\; \ell\cdot t\geqslant 2 $.}
\end{cases}
$$

\textbf{Answer:} $g_{k,h}^{({\rm{NLSFR}})} \in A({R^m})$ if
\begin{itemize}[noitemsep]
 \item $c < t$;
 \item $\ell=t=c=1,~r_c \equiv 1\,(\bmod \,2)$;
 \item $t=c,~r_c \equiv 0\,(\bmod \,2),~\ell\cdot t\geqslant 2$.
\end{itemize}

\noindent \textbf{Q2 Type-II GFS}.
Consider a permutation $\theta _h^{(2)}:{R^m} \to {R^m}$ defined by
$$\theta _h^{(2)}:({\alpha _1},...,{\alpha _m}) \mapsto ({\alpha _1},{\alpha _2} + {h_1}({\alpha _1}),{\alpha _3},{\alpha _4} + {h_2}({\alpha _3}),...,{\alpha _{m - 1}},{\alpha _m} + {h_{m/2}}({\alpha _{m - 1}}))$$
for all $({\alpha _1},...,{\alpha _m}) \in {R^m}$. It is readily seen that
$$g_{k,h}^{({\rm{GFS - II}})}({\alpha _1},...,{\alpha _m}) = \rho \delta _{{k_1}}^{(1)}\delta _{{k_2}}^{(3)}...\delta _{{k_{m/2}}}^{(m - 1)}\theta _h^{(2)}({\alpha _1},...,{\alpha _m}).$$
We have already get that if $m = {2^\ell},~\ell \geqslant 2$, then
$${\rm{sign}}\left( {\rho \delta _{{k_1}}^{(1)}\delta _{{k_2}}^{(3)}...\delta _{{k_{m/2}}}^{(m - 1)}} \right) 
 = {\rm{sign}}(\rho ) \cdot {\rm{sign}}\left( {\delta _{{k_1}}^{(1)}} \right) \cdot {\rm{sign}}\left( {\delta _{{k_2}}^{(3)}} \right) \cdot ... \cdot {\rm{sign}}\left( {\delta _{{k_{m/2}}}^{(m - 1)}} \right) = 1.$$

We now prove that ${\rm{sign}}(\theta _h^{(2)}) = 1$.
For all $i \in \{ 0,...,c\} $ and $j \in \{ 1,...,m/2\} $, we denote
$${r_i}({h_j}) = \left| {\left\{ {\beta  \in R\ |\ {h_j}(\beta ) = b,\ {\rm{ ord}}(b) = {2^i}} \right\}} \right|.$$
It is clear that $\alpha  = ({\alpha _1},...,{\alpha _m}) \in {R^m}$ belongs to a cycle of length ${2^{v(\alpha )}}$ of $\theta _h^{(2)}$, where
$$v(\alpha ) = \max \left\{ {{{\log }_2}\left( {{\rm{ord}}({h_t}({\alpha _{2t}}))} \right)\ |\ t = 1,...,m/2} \right\}.$$
For any $v \in \{ 0,...,c\}$, we define
$${U_v} = \left\{ {({j_1},...,{j_{m/2}}) \in {{\{ 0,...,c\} }^{m/2}}\ |\ v = \max \left\{ {{j_1},...,{j_{m/2}}} \right\}} \right\}.$$
The number of cycles of length ${2^v}$ is equal to
$${x_v} = {2^{t \cdot m/2 - v}}\sum\limits_{({j_1},...,{j_{m/2}}) \in {U_v}} {\prod\limits_{i = 1}^{m/2} {{r_{{j_i}}}({h_i})} } .$$
It follows that
$$\tau \left( {\theta _h^{(2)}} \right) = {2^{tm}} - \sum\limits_{v = 0}^c {{2^{t \cdot m/2 - v}}\sum\limits_{({j_1},...,{j_{m/2}}) \in {U_v}} {\prod\limits_{i = 1}^{m/2} {{r_{{j_i}}}({h_i})} } } .$$

From $t \geqslant c \geqslant v$, $m/2 \geqslant 2$, it follows that $t \cdot m/2 - v > 0$. Hence, ${x_v}$ is even for all $v \in \{ 0,...,c\} $.  Thus, $\tau \left( {\theta _h^{(2)}} \right)$ is even and ${\rm{sign}}(\theta _h^{(2)}) = 1$. Therefore,
$${\rm{sign}}\left( {g_{k,h}^{({\rm{GFS - II}})}} \right) = {\rm{sign}}(\theta _h^{(2)}) = 1.$$

\textbf{Answer:} $g_{k,h}^{({\rm{GFS - II}})} \in A({R^m})$ for all positive integers $t \geqslant 2$, $c,\ell \geqslant 1$ and mappings ${h_1},...,{h_{m/2}}$.

\noindent \textbf{Q3 TH-GFS}.
Let $\theta _h^{(3)}:{R^m} \to {R^m}$ be a mapping that for all $({\alpha _1},...,{\alpha _m}) \in {R^m}$ is such that
$$\theta _h^{(3)}:({\alpha _1},...,{\alpha _m}) \mapsto ({\alpha _1},{\alpha _2} + {h_1}({\alpha _1}),{\alpha _3} + {h_3}({\alpha _1}),...,{\alpha _m} + {h_m}({\alpha _1})).$$

It is clear that
$$g_{k,h}^{({\rm{TH}})}({\alpha _1},...,{\alpha _m}) = \rho \delta _{{k_2}}^{(2)}\delta _{{k_3}}^{(3)}...\delta _{{k_m}}^{(m)}\theta _h^{(3)}({\alpha _1},...,{\alpha _m}).$$
We have already know that if $m = {2^\ell},\ell \geqslant 2$, then
$$
{\rm{sign}}\left( {\rho \delta _{{k_2}}^{(2)}\delta _{{k_3}}^{(3)}...\delta _{{k_m}}^{(m)}} \right)
 = {\rm{sign(}}\rho {\rm{)}} \cdot {\rm{sign}}\left( {\delta _{{k_2}}^{(2)}} \right) \cdot {\rm{sign}}\left( {\delta _{{k_3}}^{(3)}} \right) \cdot ... \cdot {\rm{sign}}\left( {\delta _{{k_m}}^{(m)}} \right) = 1.
$$
Let us prove that ${\rm{sign}}(\theta _h^{(3)}) = 1$.

By ${\rm{LCM}}({a_1},...,{a_t})$ denote the least common multiple of ${a_1},...,{a_t} \in R$.

It is obvious that ${\rm{ord}}\;\theta _h^{(3)}|{2^c}$.  For each $\beta  \in R$, we define
$$w(\beta ) = {\log _2}\left( {{\rm{LCM}}\left( {{\rm{ord}}\;{h_2}{\rm{(}}\beta {\rm{)}}{\rm{,ord}}\;{h_3}{\rm{(}}\beta {\rm{)}},...,{\rm{ord}}\;{h_m}{\rm{(}}\beta {\rm{)}}} \right)} \right).$$
Let
$${r_j} = \left\{ {\alpha  \in R\ |\ w(\alpha ) = j} \right\}$$
for all $j \in \{ 0,...,c\} $.
It is clear that $({\alpha _1},...,{\alpha _m}) \in {R^m}$ belongs to a cycle of length ${2^{w({\alpha _1})}}$. The number cycles of length ${2^j}$ is equal to ${2^{t(m - 1) - j}}{r_j}$ for all $j \in \{ 0,...,c\} $.
Thus,
$$\tau \left( {\theta _h^{(3)}} \right) = {2^{tm}} - \sum\limits_{i = 0}^c {{2^{t(m - 1) - i}}{r_i}} .$$
Since $t \geqslant c \geqslant v$ and $m - 1 \geqslant 3$, we have $t \cdot (m - 1) - c > 0$. Thus, ${2^{t(m - 1) - j}}{r_j}$ is even for all $j \in \{ 0,...,c\} $. Therefore, ${\rm{sign}}(\theta _h^{(3)}) = 1$ and for all $k = ({k_2},...,{k_m}) \in {R^{m - 1}}$ it holds
$${\rm{sign}}\left( {g_{k,h}^{({\rm{TH}})}} \right) = {\rm{sign}}(\theta _h^{(3)}) = 1.$$

\textbf{Answer}: $g_{k,h}^{({\rm{TH}})} \in A({R^m})$ for all positive integers $t \geqslant 2$, $\ell,c \geqslant 1$ and mappings ${h_2},...,{h_m}$.

\subsection{Problem ``Try your quantum skills!''}

\subsubsection{Formulation}
\hypertarget{pr-ent-st}{}

In oder to use the quantum cryptanalysis techniques one should be able to work with quantum bits. Daniel knows little about quantum circuits but wants to try his hand at a new field! {\bf A quantum circuit} is a scheme where we operate with some set of qubits. The operations include one- or multi-qubit transformations provided by so called {\bf quantum gates}. They are characterized by unitary operators that act on the space of qubits.
An example of a quantum circuit is the following:
\begin{center}
\begin{quantikz}
\lstick{$\ket{0}$} & \gate{H} &\ctrl{1} &\qw& \midstick[2,brackets=none]{$\frac{1}{\sqrt{2}}(\ket{00}+\ket{11})$}\qw \\
\lstick{$\ket{0}$} & \qw & \targ{} & \qw&\qw
\end{quantikz}
\end{center}
It transforms the state $\ket{00}$ to the state $\frac{1}{\sqrt{2}}(\ket{00}+\ket{11})$. The upper wire corresponds to the action on the first qubit while the lower corresponds to the second one. Here, we have the following transformations:
\begin{center}
{\small
\noindent$\ket{00}\xrightarrow{H\text{, 1st qubit}}\frac{1}{\sqrt{2}}(\ket{0}+\ket{1}) \otimes \ket{0}\xrightarrow{CNOT\text{, both qubits}}\frac{1}{\sqrt{2}}CNOT\ket{00}+\frac{1}{\sqrt{2}}CNOT\ket{10}=\frac{1}{\sqrt{2}}(\ket{00}+\ket{11}).$
}
\end{center}
\begin{itemize}
\item[{\bf Q1}] Given the state $\ket{\psi}=\frac{1}{\sqrt{2}}(\ket{00}+\ket{11})$, design a circuit that transforms $\ket{\psi}$ to the state $\frac{1}{\sqrt{2}}(\ket{01}-\ket{10})$.

\item[{\bf Q2}] Design the circuit that distinguishes between the entangled states $\frac{1}{\sqrt{2}}(\ket{00}+\ket{11})$, $\frac{1}{\sqrt{2}}(\ket{01}+\ket{10})$ and $\frac{1}{\sqrt{2}}(\ket{01}-\ket{10})$. Distinguishing means that after the measurement of the final state we can exactly say what the state from these three was given.
Use the gates 1--5 from Table~\ref{tbl:quantum}.
\end{itemize}

\begin{remark}
\label{note-quantum} 
A qubit is a two-level quantum mechanical system whose state $\ket{\psi}$ is the superposition of basis quantum states $\ket{0}$ and $\ket{1}$. The superposition is written as $\ket{\psi}=\alpha_0\ket{0}+\alpha_1\ket{1}$, where $\alpha_0$ and $\alpha_1$ are complex numbers that possess $|\alpha_0|^2+|\alpha_1|^2=1$. The amplitudes $\alpha_0$ and $\alpha_1$ have the following physical meaning: after the measurement of a qubit which has the state $\ket{\psi}$, it will be found in the state $\ket{0}$ with probability $|\alpha_0|^2$ and in the state $\ket{1}$ with probability $|\alpha_1|^2$.
In order to operate with multi-qubit systems, we consider the bilinear operation $\otimes:\ket{x},\ket{y}\rightarrow\ket{x}\otimes\ket{y}$ on $x,y\in\{0,1\}$ which is defined on pairs $\ket{x},\ket{y}$, and by bilinearity is expanded on the space of all linear combinations of $\ket{0}$ and $\ket{1}$. When we have two qubits in states $\ket{\psi}$ and $\ket{\varphi}$ correspondingly, the state of the whole system of these two qubits is
$
\ket{\psi}\otimes\ket{\varphi}.
$
In general, for two qubits we have
$
\ket{\psi}=\alpha_{00}{\ket{0}\otimes\ket{0}}+\alpha_{01}\ket{0}\otimes\ket{1}+\alpha_{10}\ket{1}\otimes\ket{0}+\alpha_{11}\ket{1}\otimes\ket{1}.
$
The physical meaning of complex numbers $\alpha_{ij}$ is the same as for one qubit, so we have the essential restriction $|\alpha_{00}|^2+|\alpha_{01}|^2+|\alpha_{10}|^2+|\alpha_{11}|^2=1$. We use more brief notation $\ket{a}\otimes\ket{b}\equiv\ket{ab}$. For the case of multi-qubit systems with $n$ qubits the general form of the state is $
\ket{\psi}=\sum\limits_{(i_1i_2\ldots i_n)\in\{0,1\}^n}\alpha_{i_1i_2... i_n}\ket{i_1i_2\ldots i_n}.$

\end{remark}

\savebox{\boxX}{\footnotesize \begin{quantikz}
\lstick{$\ket{x}$} & \gate{X} & \qw \rstick{$\ket{x\oplus1}$}
\end{quantikz} }
\savebox{\boxZ}{\footnotesize \begin{quantikz}
\lstick{$\ket{x}$} & \gate{Z} & \qw \rstick{$(-1)^x\ket{x}$}
\end{quantikz} }
\savebox{\boxH}{\footnotesize \begin{quantikz}
\lstick{$\ket{x}$} & \gate{H} & \qw \rstick{$\frac{\ket{0}+(-1)^x\ket{1}}{\sqrt{2}}$}
\end{quantikz}}
\savebox{\boxCnot}{\footnotesize \begin{quantikz}
\lstick{$\ket{x}$} & \ctrl{1} & \qw \rstick{$\ket{x}$} \\
\lstick{$\ket{y}$} & \targ{} & \qw \rstick{$\ket{y\oplus
x}$}
\end{quantikz}}
\savebox{\boxSwap}{\footnotesize \begin{quantikz}
\lstick{$\ket{x}$} & \swap{1} & \qw \rstick{$\ket{y}$} \\
\lstick{$\ket{y}$} & \targX{} & \qw \rstick{$\ket{x}$}
\end{quantikz}}
\savebox{\boxToffoli}{\footnotesize \begin{quantikz}
                \lstick{$\ket{x}$} & \ctrl{2} & \qw \rstick{$\ket{x}$} \\
                \lstick{$\ket{y}$} & \ctrl{1} & \qw \rstick{$\ket{y}$} \\
                \lstick{$\ket{z}$} & \targ{} & \qw \rstick{$\ket{z\oplus(x\cdot y)}$}
        \end{quantikz} }

\begin{table}[h!]
{\footnotesize
\begin{tabular}{|c|p{2.3cm}|l|p{8cm}|}
\hline
1 & Pauli-X gate & \usebox\boxX & acts on a single qubit in the state $\ket{x}$, $x\in\{0,1\}$\\
\hline
2 & Pauli-Z gate & \usebox\boxZ & acts on a single qubit in the state $\ket{x}$, $x\in\{0,1\}$\\
\hline
3 & Hadamard gate & \usebox\boxH & acts on a single qubit in the state $\ket{x}$, $x\in\{0,1\}$\\
\hline
4 & controlled\,NOT (CNOT) gate & \usebox\boxCnot & acts on a pair of qubits in the states $\ket{x},\ket{y}$, $x,y\in\{0,1\}$\\
\hline
5 & SWAP gate & \usebox\boxSwap & acts on a pair of qubits in the states $\ket{x},\ket{y}$, $x,y\in\{0,1\}$\\
\hline
6 & Toffoli gate & \usebox\boxToffoli & acts on a triple of qubits in the states $\ket{x},\ket{y},\ket{z}$, $x,y,z\in\{0,1\}$\\
\hline
\end{tabular}
}
\caption{Quantum gates} 
\label{tbl:quantum}
\end{table}

\subsubsection{Solution}

\noindent{\bf Q1.} The required transformation can be described by the following circuit
	\begin{center}
		\begin{quantikz}
		\lstick[wires=2]{$\frac{1}{\sqrt{2}}(\ket{00}+\ket{11})$} & \qw & \qw & \qw\rstick[wires=2]{$\frac{1}{\sqrt{2}}(\ket{01}-\ket{10})$}\\
					\qw &\gate{Z} & \gate{X} & \qw
		\end{quantikz}
	\end{center}

\noindent{\bf Q2.} In order to distinguish between the mentioned quantum states
	$$
		\ket{\psi_1}=\frac{1}{\sqrt{2}}(\ket{00}+\ket{11}), ~~~~~
		\ket{\psi_2}=\frac{1}{\sqrt{2}}(\ket{01}+\ket{10}), ~~~~~
		\ket{\psi_3}=\frac{1}{\sqrt{2}}(\ket{01}-\ket{10})
	$$
	one can consider the following circuit:
	\begin{center}
		\begin{quantikz}
			\lstick[wires=2]{$\ket{\psi_i}$} & \ctrl{1} & \gate{H} & \qw\rstick[wires=2]{$\ket{\varphi_i}$}\\
				 & \targ{} & \qw & \qw
		\end{quantikz}
	\end{center}
Here $\ket{\varphi_i}$ denotes the output for the corresponding state. At the same time, the analysis of the output state yields the required information about the unknown input one. This distinguishing procedure comes from the results below:
\begin{center}
 for \ket{\psi_1}, we get $\ket{00}$;~~~~~
 for \ket{\psi_2}, we get $\ket{01}$;~~~~~
 for \ket{\psi_3}, we get $\ket{11}$.
\end{center}


\subsection{Problem ``Quantum error correction''}\label{sec-quantum}

\subsubsection{Formulation}
\hypertarget{pr-quantum}{}

 The procedure of error correction is required for quantum computing due to intrinsic errors in quantum gates. One of approaches to quantum error correction is to encode quantum information in three-qubit states, i.\,e. $\alpha_0\ket{0}+\alpha_1\ket{1}\rightarrow\alpha_0\ket{000}+\alpha_1\ket{111}$. 
 
 Below are \underline{\bf Problems for a special prize!}
 \begin{itemize}
     \item[{\bf Q1}] Design a circuit which implements such encoding.

     \item[{\bf Q2}] Design a circuit which restores the initial state of the three-qubit system, if a single bit-flip error $\ket{0}\leftrightarrow\ket{1}$ occurs in one of three qubits.
      Hint: use two additional qubits and three-qubit Toffoli~gates.

     \item[{\bf Q3}] What will happen, if the quantum gates used for error correction are imperfect? What will be the threshold for gate fidelity, when the error correction will stop working?
 \end{itemize}

\begin{remark}
Please use the basic information from the Remark\;\ref{note-quantum} and gates from Table\;\ref{tbl:quantum}.
\end{remark}

\subsubsection{Solution}
		
{\bf Q1}. The encoding can be described by the following circuit:
			\begin{center}
				\begin{quantikz}
					\lstick{$\alpha\ket{0}+\beta\ket{1}$} & \ctrl{1} & \qw & \qw\rstick[wires=3]{$\alpha\ket{000}+\beta\ket{111}$} \\ 
					\lstick{$\ket{0}$} & \targ{} & \ctrl{1} & \qw \\
					\lstick{$\ket{0}$} & \qw & \targ{} & \qw
				\end{quantikz}
			\end{center}

\noindent{\bf Q2}. Let us firstly describe the authors' solution. To find the bit-flip in each qubit, we introduce two ancillary qubits and entangle them with our three data qubits via CNOT gates:
		\begin{center}
		    \begin{quantikz}
				\lstick[wires=3]{$\alpha\ket{000}+\beta\ket{111}$} & \gate[3,style={starburst,fill=yellow,draw=red,line
					width=2pt,inner xsep=-4pt,inner ysep=-5pt},
				label style=cyan]{\text{bit-flip}} & \qw & \ctrl{3} & \qw & \ctrl{4} & \qw & \qw \\ 
				& & \qw & \qw & \ctrl{2} & \qw & \qw & \qw \\
				& & \qw & \qw & \qw & \qw & \ctrl{2} & \qw \\
				\lstick{$\ket{0}$} & \qw & \qw & \targ{} & \targ{} & \qw & \qw & \qw\\
				\lstick{$\ket{0}$} & \qw & \qw & \qw & \qw & \targ{} & \targ{} & \qw
			\end{quantikz}
		
		\end{center}
	
		Without bit-flips in the data qubits, both ancillary qubits will stay in the state $\ket{00}$, because the states of data qubits are identical. It means that, depending on the initial state of the first qubit, the Pauli-X gate will be either never applied to the ancillary qubits, or applied twice.
		
		If there is a bit-flip in any of data qubits, the Pauli-X gate will be applied once or three times to one of the ancillary qubits. This will indicate the error in the particular data qubit:\\
		\smallskip
		    \indent $\bullet$ state $\ket{00}$ means ``no error'';\\
		    \indent $\bullet$ state $\ket{11}$ means ``error in the 1st qubit'';\\
		 	\indent $\bullet$ state $\ket{10}$ means ``error in the 2nd qubit'';\\
			\indent $\bullet$ state $\ket{01}$ means ``error in the 3d qubit''.
\smallskip
		
		Now it is possible to restore the initial state by applying Toffoli gates. For example, a Toffoli gate with two ancillary qubits used as control ones and first data qubit used as target ones will flip its state if the ancillary qubits are in state and leave it unchanged in any other case (no error in the first qubit). Similarly, the flips in other qubits can be restored. The final circuit is
		\begin{center}
			\begin{quantikz}
				\lstick[wires=3]{$\alpha\ket{000}+\beta\ket{111}$} & \gate[3,style={starburst,fill=yellow,draw=red,line
					width=2pt,inner xsep=-4pt,inner ysep=-5pt},
				label style=cyan]{\text{bit-flip}} & \qw & \ctrl{3} & \qw & \ctrl{4} & \qw\slice{} & \targ{} & \qw & \qw & \qw & \qw & \qw & \qw \\ 
				& & \qw & \qw & \ctrl{2} & \qw & \qw & \qw & \qw & \targ{} & \qw & \qw & \qw & \qw\\
				& & \qw & \qw & \qw & \qw & \ctrl{2} & \qw & \qw & \qw & \qw & \targ{} & \qw & \qw\\
				\lstick{$\ket{0}$} & \qw & \qw & \targ{} & \targ{} & \qw & \qw & \ctrl{-3} & \qw & \ctrl{-2} & \targ{} & \ctrl{-1} & \targ{} & \qw\\
				\lstick{$\ket{0}$} & \qw & \qw & \qw & \qw & \targ{} & \targ{} & \ctrl{-4} & \targ{} & \ctrl{-3} & \targ{} & \ctrl{-2} & \qw & \qw
			\end{quantikz}
		\end{center}

		During the Olympiad, twelve teams made progress in solving the problem and suggested good and correct schemes. We would like to mention the best one proposed by the team of Viet-Sang Nguyen, Nhat Linh Le Tan, Nhat Huyen Tran Ngoc (France, Paris). Taking into account discussions on {\bf Q3} in the solution of this team, we mark this problem as ``partially solved''. In their circuit, only one Toffoli gate is used:
		\begin{center}
			\begin{quantikz}
				\lstick[wires=3]{$\alpha\ket{000}+\beta\ket{111}$} & \gate[3,style={starburst,fill=yellow,draw=red,line
					width=2pt,inner xsep=-4pt,inner ysep=-5pt},
				label style=cyan]{\text{bit-flip}} & \qw & \ctrl{3}\gategroup[5,steps=9,style={dashed,
					rounded corners,fill=blue!10, inner xsep=2pt},
				background]{{Error-correction stage}} & \targ{} & \qw & \qw & \qw & \qw & \qw & \qw & \targ{} & \qw\\ 
				& & \qw & \qw & \qw & \ctrl{3} & \targ{} & \qw & \qw & \qw & \targ{} & \qw & \qw \\
				& & \qw & \qw & \qw & \qw & \qw & \ctrl{1} & \ctrl{2} & \targ{} & \ctrl{-1} & \ctrl{-2} & \qw \\
				\lstick{$\ket{0}$} & \qw & \qw & \targ{} & \ctrl{-3} & \qw & \qw & \targ{} & \qw & \ctrl{-1} & \qw & \qw & \qw\\
				\lstick{$\ket{0}$} & \qw & \qw & \qw & \qw & \targ{} & \ctrl{-3} & \qw & \targ{} & \ctrl{-2} & \qw & \qw & \qw
			\end{quantikz}
		\end{center}

\noindent {\bf Q3.} Several participants proposed interesting ideas on this problem.
		In some of them, the minimum fidelities for a success probability were considered independently for every type of gates, i.\,e. Pauli-X, CNOT and Toffoli gate, and corresponding diagrams were shown. In another, it was assumed that the probability of imperfect operation of each gate is the same, then the threshold when error correction stops working was estimated. 
		
		There was an approach under assumption that the error-box makes a single bit-flip error and the error-correction box makes a mistake, both with some fixed probabilities, and the probability that the error-box makes multiple bit-flip errors is neglectable. It was obtained that the error-correction stops working when the probability of its proper is larger than $1/2$.

\subsection{Problem ``$s$-Boolean sharing''}\label{sec-sharing}

\subsubsection{Formulation}
\hypertarget{pr-sB-sharing}{}

In cryptography, a field known as {\bf side-channel analysis} uses extra information such as the power consumption of an implementation to break a cryptographic primitive. In order to defend against these attacks, one does not need to change the primitive but only the way the primitive is implemented. A popular countermeasure is called ``{\bf sharing}'' where the computation of the primitive is split in multiple parts (this notion was firstly suggested in \cite{99-Chari, 99-Goubin}). Each part seemingly operates on random data such that an adversary has to observe all parts of the computation in order to gain sense of the secret information that was processed.

 An $s$-{\bf Boolean sharing of a variable} $x \in \mathbb{F}_2$ is a vector $(x_1, x_2, ..., x_{s})\in\mathbb{F}_2^s$ such that $x = \bigoplus_{i = 1}^{s} x_i$.
A vectorial Boolean function $G:\mathbb{F}_2^{sn} \rightarrow \mathbb{F}_2^{sm}$ is an $s$-{\bf Boolean sharing of a function} $F:\mathbb{F}_2^n \rightarrow \mathbb{F}_2^m$ if for all $x \in \mathbb{F}_2^n$ and $(x_1,...,x_s) \in \mathbb{F}_2^{sn}$, $x_i\in\mathbb{F}_2^n$, such that $\bigoplus_{i=1}^s x_i = x$, $$\bigoplus_{i=1}^s G_i(x_1,...,x_s) = F(x)\,.$$
Here, $G = (G_1,...,G_s)$, where $G_i:\mathbb{F}_2^{sn} \rightarrow \mathbb{F}_2^{m}$  and ``$\oplus$'' denotes the bit-wise XOR.

\begin{itemize}
    \item[{\bf Q1}]
Write an algorithm which takes in a vectorial Boolean function and an integer $s$ and returns true/false on whether the function is a $s$-Boolean sharing of another function. In case the result is true, the algorithm also returns the function whose sharing is the algorithm's input. 
\item[{\bf Q2}] \underline{\bf Problem for a special prize!} Propose a theoretical solution to the problem of checking whether the function is a $s$-Boolean sharing of another function.
\end{itemize}

\noindent {\bf Example.} If you give the Boolean function $G: \mathbb{F}_2^6 \rightarrow \mathbb{F}_2^3$ such that
\begin{align*}
    G_1(a,b,c,d,e,f) &= ad \oplus ae \oplus bd \\
    G_2(a,b,c,d,e,f) &= be \oplus bf \oplus ce \\
    G_3(a,b,c,d,e,f) &= cf \oplus cd \oplus af
\end{align*}
the algorithm should return true when $s=3$ together with the function $F: \mathbb{F}^2_2 \rightarrow \mathbb{F}_2$ such that $F(x,y) = xy$, where $x = a \oplus b \oplus c$ and $y = d \oplus e \oplus f$.

\subsubsection{Solution}

{\bf Solution to Q1.} We will give a general approach.
Consider a function $G:\mathbb{F}_2^{sn} \rightarrow \mathbb{F}_2^{sm}$ of variables $x_1,...,x_{sn}$, we check whether it is an $s$-Boolean sharing of some function $F:\mathbb{F}_2^n \rightarrow \mathbb{F}_2^m$. Take an arbitrary permutation of the $sn$ input bits $\pi$, there are a total of $sn!$ of such permutations (we note that one can reduce this number as some permutations would lead to the same sharing). Denote $\pi(x_1,...,x_{sn}) = (y_1,...,y_{sn})$ and $z_i = (y_{(i-1)*n+1},...,y_{i*n})$ for $i \in \{1,...,s\}$. We want to verify whether 
$$\bigoplus_{i=1}^{s} G_i(z_1,...,z_{s}) = F( \bigoplus_{i=1}^{s} z_i)\,,$$
for all $(z_1,...,z_{s}) \in \mathbb{F}_2^{sn}$.
This is easily done via a brute force approach of going through all $(z_1,...,z_{s}) \in \mathbb{F}_2^{sn}$ (this requires $2^{sn}$ evaluations) and verifying the above equation. In case the equation does not hold, we go to the next permutation $\pi$. Otherwise, we stop searching and return true. The algorithm would require around $sn! \cdot 2^{sn}$ steps.

{\bf Ideas on Q2.}
The most interesting idea found by the participants considers the algebraic normal form of the shared function.
Let us consider an ordered case where it is known which inputs would form the shares of the function. Let $F$ be an arbitrary Boolean function. In case $F$ is the unshared function of some $G$, then 
$$\bigoplus_{i=1}^{s} G_i(x_1,...,x_{s}) = F( \bigoplus_{i=1}^{s} x_i)\,,$$
Notice that for each monomial $x^1\cdot ... \cdot x^\ell$ in $F$, we get the shared monomial $(\bigoplus_i x_i^1)\cdot ... \cdot (\bigoplus_i x_i^\ell)$. We then verify for each monomial in $G$ whether the other shares of that monomial are also present. If so, we remove $(\bigoplus_i x_i^1)\cdot ... \cdot (\bigoplus_i x_i^\ell)$ and repeat until no more monomial are present in $G$.

The best solution found was given by the team of university students Gongyu Shi, Ruoyi Kong, Haoxiang Jin (China, Shanghai) and awarded a special prize for ``partially solving'' the problem.

\subsection{Problem ``Let's find permutations!''}

\subsubsection{Formulation}
\hypertarget{pr-APN+lin-perm}{}

A function $F$ from $\mathbb{F}_{2^n}$ to itself is called {\bf APN} ({\bf almost perfect nonlinear}) if for  any $a,b \in \mathbb{F}_{2^n}$ with $a \ne 0$ the equation $F(x) + F(a+x) = b$  has at most 2 solutions. APN functions possess an optimum resistance to differential cryptanalysis and are under the extreme interest in cryptography! For example, when the unique 1-to-1 APN function in 6 variables was found in 2009, it was immediately applied in construction of the known lightweight cipher FIDES.

Let $F(x)=x^d$. It is known that $F$ is APN for the following exponents $d$:
\begin{itemize}[noitemsep]
\item $d=2^{2i}-2^i+1$, ${\rm gcd}(i,n)=1$, $2\leqslant i\leqslant n/2$;
\item $d=2^t+3$, $n=2t+1$;
\item $d=2^t+2^{t/2}-1$ for  $t$ even and $d=2^t+2^{(3t+1)/2}-1$ for  $t$ odd with $n=2t+1$;
\item $d=2^{2t}-1$, $n=2t+1$;
\item $d=2^{4i}+2^{3i}+2^{2i}+2^{i}-1$ with $n=5i$.
\end{itemize}

\begin{itemize}
\item[{\bf Q1}] \underline{\bf Problem for a special prize!} Describe (characterize or make a list of) all linear functions $L_1$ and $L_2$ for any one exponent above for $n=7$ or $n=8$, such that the function $L_1(x)+L_2(F(x))$ is a permutation.

\item[{\bf Q2}] \underline{\bf Problem for a special prize!}
Consider any of the exponents $d$ above. Find linear functions $L_1$ and $L_2$ (both different from 0 function) such that the function $L_1(x)+L_2(F(x))$ is a permutation ($n\geqslant9$), or prove that such functions do not exist.
\end{itemize}

\begin{remark}
 $\mathbb{F}_{2^n}$ is the finite field of order $2^n$. A function $F:\mathbb{F}_{2^n}\to\mathbb{F}_{2^n}$  has the unique representation $F(x) = \sum_{i=0}^{2^n-1} c_i x^{i}, c_i\in\mathbb{F}_{2^n}$.
The algebraic degree of $F$ is equal to the maximum binary weight of $i$ such that $c_i\neq 0$.
A linear function $L$ has degree at most 1 and $L(0)=0$ (that is $L(x) = \sum_{k=1}^{n} c_k x^{2^k}$).
\end{remark}

\subsubsection{Solution}

The problems discussed are related to the problem of relation between CCZ- and EA-equivalences for power APN functions. This was studied in \cite{20-CCZ-EA}.
Regarding {\bf Q1}, the problem is solved for $n\leqslant 9$ in \cite{20-Calderini} in terms of codes. The  only possible cases are the following:
\begin{itemize}[noitemsep]
    \item  for $n=7$, $L_1=0$ and $L_2$ is a permutation, or $L_2=0$ and $L_1$ is a permutation;
    \item for $n=8$, $L_2=0$ and $L_1$ is a permutation.
\end{itemize}

Regarding {\bf Q2}, there were no great ideas proposed by the participants. The one nontrivial solution was given by Alexey Chilikov (Russia, Moscow).

\subsection{Problem ``Distance to affine functions''}

\subsubsection{Formulation}
\hypertarget{pr-distance}{}

Given two functions $F$ and $G$ from $\mathbb{F}_2^n$ (or $\mathbb{F}_{2^n}$) to itself, their Hamming distance equals by definition the number of inputs $x$ at which $F(x)\neq G(x)$. The minimum Hamming distance between any such function $F$ and all affine functions $A$  is known to be strictly smaller than $2^n-n-1$ if $n\geqslant 4$.


Consider the following problems. Each of them is a \underline{\bf Problem for a special prize!}

\begin{itemize}
\item[{\bf Q1}]
Find a better upper bound valid for every $n$.

\item[{\bf Q2}] If {\bf Q1} is unsuccessful, find constructions of infinite classes of functions $F$ having a distance to affine functions as large as possible (infinite classes meaning that these functions are in numbers of variables ranging in an infinite set, such as all positive integers, possibly of some parity for instance).

\item[{\bf Q3}]
 If {\bf Q1} and {\bf Q2} are unsuccessful, find constructions (possibly with a computer; then a representation of these functions will be needed, such as their algebraic normal form or their univariate representation) of functions $F$ in fixed numbers of variables having a distance to affine functions as large as possible.
\end{itemize}

\begin{remark} We recall that an affine function $A$ is a function satisfying $A(x)+A(y)+A(z)=A(x+y+z)$ for all inputs $x,y,z$.
\end{remark}

\subsubsection{Solution}

The bound $<2^n-n-1$ if $n\geqslant 4$ was found in \cite[Section 7]{21-Carlet}. The problems discussed are connected with a curious open problem of finding bounds on the nonlinearity of differentially uniform functions.

The most interesting ideas were presented by the team of Gabor P. Nagy, Gabor V. Nagy, and  Miklos Maroti (Hungary, Budapest) and concerned the relation with differential uniformity of a special function.


\end{document}